\documentstyle [preprint, aps]{revtex}
\renewcommand{\thefootnote}{\alph{footnote}}
\renewcommand{\thesection}{\Roman{section}}
\renewcommand{\theequation}{\thesection.\arabic{equation}}

\begin{document}
\renewcommand{\thefootnote}{\alph{footnote}}
\title{
\renewcommand{\thefootnote}{\alph{footnote}}
Chiral limit of the two-dimensional fermionic determinant
in a general magnetic field}
\author{M. P. Fry\renewcommand{\thefootnote}{\alph{footnote}}
\footnote{\normalsize Electronic mail: \tt {mpfry@maths.tcd.ie}}}        
\address{\normalsize\it School of Mathematics, University of Dublin, \\
         \normalsize \it Dublin 2, Ireland}
\maketitle
\begin{abstract}
We consider the effective action for massive two-dimensional QED in flat Euclidean space-time in the background of a general square-integrable magnetic field 
with finite range. It is shown that its small mass limit is controlled by the 
chiral anomaly. New results for the low-energy scattering of electrons in 2+1 dimensions in static, inhomogenous magnetic 
fields are also presented. 
\end{abstract}
\newpage
\section{INTRODUCTION}
\setcounter{equation}{0}
\renewcommand{\thesection}{\arabic{section}}
\renewcommand{\theequation}{\thesection.\arabic{equation}}

Fermionic determinants lie at the heart of gauge field theories with fermions. They are obtained by integrating over the fermionic degrees of freedom in the presence of a background potential $A_{\mu}$, producing the one-loop effective action $S_{\mbox{{\scriptsize{eff}}}}= - \ln \mbox{det}$, where the fermionic determinant, det, is formally the ratio 
$\mbox{det}(\not\! P - \not\!\! A + m)/ 
\mbox{det}( \not\! P + m)$
of determinants of Dirac operators. 
The coupling constant $e$ has been absorbed into $A_{\mu}$.
This action is exact and appears in the calculation of every physical process.
Therefore, any truly nonperturbative calculation must deal with 
$S_{\mbox{{\scriptsize{eff}}}}$ in its full generality. The main problem with calculating  $S_{\mbox{{\scriptsize{eff}}}}$ is that it must be known for generic potentials, typically tempered distributions, if it is to be part of an effective measure for $A_{\mu}$.
A summary of what is known about $S_{\mbox{{\scriptsize{eff}}}}$ in quantum electrodynamics in 1+1, 
2+1 and 3+1 dimensions for general fields is given in Sec.I of Ref.1. 
Recall that  $S_{\mbox{{\scriptsize{eff}}}}$ in QED only depends on the field strength tensor 
$F_{\mu\,\nu}$.
It is seen that there are upper and lower bounds on $S_{\mbox{{\scriptsize{eff}}}}$, with some bounds holding only for restricted fields, such as unidirectional ones. After fifty years or so there are still no equalities in QED for general fields, except for massless QED in 1+1 dimensions - the Schwinger 
model~\cite{ii}.
\paragraph*{}
In this paper an equality is obtained for the chiral limit of  $S_{\mbox{{\scriptsize{eff}}}}$ in two-dimensional, Wick-rotated Euclidean QED for a general field, hereafter referred to as a static magnetic field $B(\bf{r})$. 
Of course $B$ is not completely unrestricted. We described 
elsewhere~\cite{i,v} precisely, how rough potentials and fields are to be smoothed as part of the regularization process required to make the functional
integration over $A_{\mu}$ well-defined. It is sufficient to assume in this paper that $A_{\mu}$ is differentiable and that $B$ is square integrable with finite range $R$. Then $B$ is guaranteed to have finite flux, $\Phi$, since 
$||B|| \ge |\Phi|/ \sqrt \pi \, R$, where $||B||^2 = \int d^2r B(\bf{r})^2$ 
and $\Phi=\int d^2r B(\bf{r})$.
The author knows of no definition of a determinant that can handle infinite flux fields; there is simply too much degeneracy~\cite{vi}, resulting in volume-like divergences (which are ignored) as in the constant field case. Furthermore, finite flux and range are consistent with the need to introduce a volume cutoff to define  
$\mbox{QED}_2$ before taking the thermodynamic limit.

With the foregoing restrctions on $B$ our result is
\begin{equation}
\lim_{m^2 \to 0} \,m^2 \frac{\partial}{\partial m^2} \ln \mbox{det}
=
\frac{|\Phi|}{4 \pi}\,,
\label{eqn:ione}
\end{equation}
where $m$ is the fermion mass. Together with the exact scaling relation
\begin{equation}
\ln \mbox{det}(\lambda^2 B(\lambda {\bf r}), m^2) = 
\ln \mbox{det}(B({\bf r}), m^2/\lambda^2)\,,
\label{eqn:itwo}
\end{equation}
(\ref{eqn:ione}) implies the strong field limit
\begin{equation}
\ln \mbox{det}(\lambda^2 B(\lambda {\bf r}), m^2) 
\lower5pt \hbox{$\scriptscriptstyle \lambda \gg 1$}
\mkern-25mu \sim
- \frac{|\Phi|}{2 \pi} \ln \lambda + R(\lambda)\,,
\label{eqn:ithree}
\end{equation}
where 
$
\lower5pt \hbox{$\scriptscriptstyle \lambda\to \infty $}
\mkern-30mu \lim (R(\lambda)/ \ln \lambda) = 0\,.
$
Note that the chiral limit in (\ref{eqn:ione}) implies that 
$\mbox{QED}_2$'s fermionic determinant behaves like 
$( |\Phi|/4 \pi ) \ln m^2$ as $m \rightarrow 0$,
which does not coincide with that of the Schwinger model. 
\paragraph*{}
For nonwinding background fields with $\Phi =0$ one can prove continuity at 
$m=0$. As a result, massive $\mbox{QED}_2$'s fermionic determinant does coincide with that of the Schwinger model at $m=0$:
\begin{equation}
\lim_{m\rightarrow 0} \ln \mbox{det}=
\frac{1}{4 \pi^2}
\int d^2r d^2r^{\prime} B({\bf r}) B({\bf r^{\prime}}) 
\ln|{\bf r - r^{\prime}}|\,.
\end{equation}
This follows from results of Seiler~\cite{iii} and Simon~\cite{iii.i} as will be 
shown in a future paper.
\paragraph*{}
It is reasonable to ask what is the relevance of $\mbox{QED}_2$'s fermionic 
determinant, and its mass dependence in particular, to physics? The answer 
is that the integral of this determinant over the fermion mass fully 
determines  $\mbox{QED}_4$'s fermionic determinant for the same magnetic field 
$B({\bf r})$~\cite{iii.ii}.
This determinant is still unknown except for a constant field~\cite{heis,vii} and
a $\mathrm{sech}^2(x/R)$ varying unidirectional field~\cite{iii.iii}.
\paragraph*{}
From the input parameters to the $\mbox{QED}_2$ determinant one can form the 
dimensionless ratios $e ||B||/m c^2$ and $\hbar/mc R$, where $e$ has been temporarily restored.
This paper deals with the nonperturbative, small mass region $e ||B||/m c^2$, 
$\hbar/mc R\gg 1$. The large mass region can be dealt with by a derivative expansion of $\ln \mbox{det}$~\cite{iii.iv}. What remains in order to estimate $\mbox{QED}_4$'s
 fermionic determinant for general unidirectional fields are optimal upper and lower bounds on $\mbox{QED}_2$'s determinant for intermediate values of the mass.
\paragraph*{}
The derivation of (\ref{eqn:ione}) is really just a problem in quantum mechanics dealing with a particle confined to a planar surface with an inhomogenous magnetic field normal to it. The proportionality of the limit in 
(\ref{eqn:ione}) to the two-dimensional chiral anomaly, $\Phi/2 \pi$, as well as its sign and its connection with paramagnetism, are discussed in Ref.12.
Equation~(\ref{eqn:ione}) was established in Ref.12 in finite volume for a unidirectional field $B({\bf r}) \ge 0$. There is a missing volume 
factor in 
Equation~(2.7) that was corrected in Ref.13. These restrictions are dropped in this paper.
\paragraph*{}
Finally, the chiral limit of $\mbox{QED}_2$'s continuum fermionic determinant should provide a nontrivial test of algorithms for the determinant on large 
lattices.
The reason is that chiral limits and topological invariants - the chiral anomaly in this instance - are notoriously difficult to calculate on a 
lattice~\cite{vii.i}.
Many of the results here on low-energy scattering in static, inhomogenous magnetic fields are new and are relevant to the physical case of electrons in such fields in 2+1 dimensions.
\paragraph*{}
In Sec.II we discuss how we will demonstrate (\ref{eqn:ione}). Section III develops the essentials of low-energy scattering in inhomogenous magnetic fields that will be required. In Sec.IV the crucial argument that central symmetry is sufficient to establish (\ref{eqn:ione}) is given. Finally, Sec.V gives the fine points of the limit (\ref{eqn:ione}).
\renewcommand{\thesection}{\Roman{section}}
\renewcommand{\theequation}{\thesection.\arabic{equation}}
\section{PRELIMINARIES}
\setcounter{equation}{0}
\renewcommand{\thesection}{\arabic{section}}
\renewcommand{\theequation}{\thesection.\arabic{equation}}

We adopt Schwinger's proper time definition~\cite{vii} of the fermionic 
determinant for Euclidean $\mbox{QED}_2$:
\begin{equation}
\ln \mbox{det} =
\frac{1}{2} \int_{0}^{\infty} \,\frac{dt}{t}\, \mbox{Tr }( e^{- P^2 t} 
- exp\{-[({\bf P} - {\bf A})^2 - \sigma_3 B] t\} ) e^{-t m^2}\,.
\label{eqn:iione}
\end{equation}
Then 
\begin{equation}
\frac{\partial}{\partial m^2} \ln \mbox{det} =
\frac{1}{2} \mbox{Tr } [ (D^2 -  \sigma_3 B + m^2)^{-1} - (P^2 + m^2)^{-1} ]\,,
\label{eqn:iitwo}
\end{equation}
where $D^2=( {\bf P} - {\bf A})^2$  and $\sigma_3$ is the Pauli matrix.
Now introduce the sum rule~\cite{viii}
\begin{equation}
 \mbox{Tr }  
[ (D^2 - B + m^2)^{-1} - (D^2 +B + m^2)^{-1} ]
=
\frac{\Phi}{2 \pi m^2}\,,
\label{eqn:iithree}
\end{equation}
where the trace is over space indices only, and assume without loss of generality that $\Phi > 0$. Then (\ref{eqn:iitwo}) and (\ref{eqn:iithree}) give
\begin{equation}
m^2  \frac{\partial}{\partial m^2} \ln \mbox{det} =
\frac{\Phi}{4 \pi} + m^2 
\mbox{Tr }  
[ (D^2 + B + m^2)^{-1} - (P^2 + m^2)^{-1} ]\,.
\label{eqn:iifour}
\end{equation}
The continuum part of the spectrum of the negative chirality operator $D^2 +B$
stretches down to zero in the case of open spaces. Because $\Phi > 0$ the
square-integrable zero modes are confined to the spectrum of  
$D^2 -B$~\cite{ix}. Although $B$ has no definite sign, its flux does, and 
 $\Phi$, chirality and the number of  square-integrable zero modes of the 
supersymmetric pair of operators $D^2 \pm B$ are correlated by the Aharonov
-Casher theorem. At this stage the minor modifications one has to make to 
deal with the case when $\Phi < 0$ are already clear.
\paragraph*{} 
It might seem that (\ref{eqn:iifour}) makes 
(\ref{eqn:ione}) self-evident. But if (\ref{eqn:iithree}) is multiplied by $m^2$ and the limit $m^2=0$ taken, then the fractional part of the chiral anomaly
is given by a difference of zero-energy phase shifts of opposite chirality~\cite{x}, demonstrating that the trace difference in (\ref{eqn:iithree})
develops a $1/m^2$- type singularity at the bottom of the continuum.
How, then, does one know a priori that such a singularity is absent from the
trace in (\ref{eqn:iifour})?
\paragraph*{}
Our definition of the determinant in (\ref{eqn:iione}) leads us to define the trace in  (\ref{eqn:iifour}) by a difference of diagonal heat kernels,
\begin{equation}
\mbox{Tr }  
[ (D^2 + B + m^2)^{-1} - (P^2 + m^2)^{-1} ]
=
\int_{0}^{\infty} \, dt e^{- t m^2} \int\,d^2r 
<\!{\bf r}\,|\,e^{-(D^2 + B)t} - e^{-P^2 t}\,|\,{\bf r}\!>\,.
\label{eqn:iifive}
\end{equation}
Denote the scattering states of $D^2 + B$ corresponding to outgoing radial 
waves by $\psi^{(+)}({\bf k},{\bf r}) = <\!{\bf r}\,|\,{\bf k},\mbox{in}\!>$ whose eigenvalues
are $E = k^2$. These satisfy the normalization condition
\begin{equation}
\int\,d^2r \,\psi^{(+)\,\ast}({\bf k},{\bf r}) \,
\psi^{(+)}({\bf k^{\prime}},{\bf r})
=
\delta({\bf k} - {\bf k^{\prime}})\,.
\label{eqn:iisix}
\end{equation}
Assume that $B({\bf r})$ is noncentral. Let $\Theta$ denote the direction of the incident beam with momentum ${\bf k}$ relative to an axis fixed in the scattering center. The asymptotic behavior of  $\psi^{(+)}({\bf k},{\bf r})$ for 
$ k r \gg 1$ is
\begin{equation}
\psi^{(+)}({\bf k},{\bf r}) =
\frac{1}{2 \pi}\, e^{i k r \, \cos(\theta - \Theta)} + \frac{f(\theta,\Theta)}
{2 \pi \sqrt r} e^{i k r} + R\,,
\label{eqn:iiseven}
\end{equation}
where $f$ is the scattering amplitude and $R$ is the remainder in the large-$r$
expansion of $\psi^{(+)}$. Equation (\ref{eqn:iiseven}) is obtained from the 
Lippmann-Schwinger equation
\begin{equation}
\psi^{(+)}({\bf k},{\bf r}) =
\frac{1}{2 \pi}\, e^{i {\bf k}\cdot {\bf r}} -
\frac{i}{4} \int \,d^3 r^{\prime} \, H_0^{(+)}(k |{\bf r - r^{\prime}}|) 
V({\bf r^{\prime}}) \psi^{(+)}({\bf k},{\bf r^{\prime}})\,,
\label{eqn:iieight}
\end{equation} 
where $ H_0^{(+)}$ is a Hankel function of the first kind and
\begin{equation}
V= -{\bf P \cdot A - A \cdot P} + A^2 + B\,.
\label{eqn:iinine}
\end{equation}
As we will show later, we can choose a gauge such that
\begin{equation}
{\bf A}=
\frac{\Phi}{2 \pi r} \,{\mathbf \hat{\theta}}\,,
\label{eqn:iiten}
\end{equation}
for $r$ sufficiently large, where ${\mathbf\hat{\theta}}$ is a unit vector orthogonal to ${\bf r}$.
Therefore, we are dealing with a long range ($1/r^2$) potential $V$,
and this is what makes the proof of (\ref{eqn:ione}) nonroutine.
The completeness of the ``in'' states for $D^2 + B$ and (\ref{eqn:iifive})
give
\begin{eqnarray}
&\mbox{Tr }&  
[ (D^2 + B + m^2)^{-1} - (P^2 + m^2)^{-1} ] 
\nonumber \\
&=& \int_{0}^{\infty} \, dt e^{- t m^2} \int\,d^2r \,
\int_0^{\infty}\,dk\,k e^{-k^2 t}
\int_{0}^{2 \pi}\,d\Theta 
(|\psi^{(+)}({\bf k},{\bf r})|^2 -
|\psi_{0}({\bf k},{\bf r})|^2)   \,,
\label{eqn:iieleven}
\end{eqnarray}
where $ \psi_{0}({\bf k},{\bf r}) = e^{ i {\bf k \cdot r}}/2 \pi$.
We are interested in the small $m^2$, high $t$ limit of (\ref{eqn:iieleven}) 
which is determined by the low energy end of the spectrum of $D^2 + B$.
Therefore we cut off the energy integral in (\ref{eqn:iieleven}) at
$M$ with $M R\ll 1$ and consider, for $m^2 \rightarrow 0$,
 \begin{eqnarray}
&\mbox{Tr }&  
[ (D^2 + B + m^2)^{-1} - (P^2 + m^2)^{-1} ] 
\nonumber \\
&=&\int_{0}^{\infty} \, dt e^{- t m^2} \int\,d^2r \,
\int_0^{M}\,dk\,k e^{-k^2 t}
\int_{0}^{2 \pi}\,d\Theta 
(|\psi^{(+)}({\bf k},{\bf r})|^2 -
|\psi_{0}({\bf k},{\bf r})|^2) + R(m^2) \,.
\label{eqn:iitwelve}
\end{eqnarray}
The remainder, $R$, can be put in the form
$$
\int_0^{\infty} e^{- t (m^2 + M^2)}
\int d^2r 
\int_0^{\infty}dp p e^{-p^2 t}
\int_{0}^{2 \pi} d\Theta 
(|\psi^{(+)}(\sqrt{p^2 + M^2},{\bf r},\Theta)|^2 -
|\psi_{0}(\sqrt{p^2 + M^2},{\bf r},\Theta)|^2) \,,
$$
which makes the energy gap between 0 and $M$ evident so that
$
\lower8pt \hbox{$\scriptscriptstyle m^2 \to 0 $}
\mkern-30mu \lim m^2 R(m^2) = 0\,.
$
Thus, (\ref{eqn:iifour}) shows that (\ref{eqn:ione}) will be established if the integral in (\ref{eqn:iitwelve}) multiplied by $m^2$ vanishes in the limit 
$m^2=0$. 
\paragraph*{}
We will calculate in the Lorentz gauge $\partial_{\mu} A_{\mu} = 0$
which, in two dimensions, allows us to set
$A_{\mu} = \epsilon_{\mu\nu} \partial_{\nu} \phi$ with
\begin{equation}
B(r, \theta)
= - \partial^2 \phi(r, \theta)\,,
\label{eqn:iififteen}
\end{equation}
and $\epsilon_{\mu\nu} = - \epsilon_{\nu\mu}$ with $\epsilon_{12}=1$.
Assuming that $B$ has range $R$ we can calculate $\phi$ in a disk $D$ of radius
 $a\gg R$ with $\phi(a, \theta)=0$.
A unique solution of Poisson's equation with Dirichlet boundary conditions
requires that we also specify $\phi$ as $r \rightarrow \infty$, which we will do by requiring that $\phi$ approach the potential of a flux line through the origin. The construction of the Dirichlet Green's function for this problem is standard, with the result
\begin{eqnarray}
\phi(r, \theta) 
& = &
-\frac{1}{4 \pi} \int_{D} d^2 r^{\prime} B(r^{\prime}, \theta^{\prime})
\ln 
\left(
\frac{r^2 + r^{\prime}{}^2 - 2 r r^{\prime} \cos ( \theta^{\prime} - \theta)}
{a^2 + \frac{r^2 r^{\prime}{}^2}{a^2}  - 
2 r r^{\prime} \cos ( \theta^{\prime} - \theta)} 
\right)\,,\;\;r < a
\nonumber \\
& = &
-\frac{\Phi}{ 2 \pi} \ln(r/a)\,,\;\; r > a\,.
\label{eqn:iisixteen}
\end{eqnarray}
This potential results in a discontinuity in $\partial \phi/\partial r$ at
$r =a$ that is of order $||B|| R^2/a^2$ and which vanishes in the limit of 
radial 
symmetry. This introduces a zero-flux, zero-range magnetic field on a ring 
at $r = a$ that does not affect the low energy phase shifts.
What is gained by this is radial symmetry for $r > a$.
\paragraph*{}
From  (\ref{eqn:iitwelve}) it is evident that we will need the outgoing wave solution of
\begin{equation}
[ ({\bf P - A})^2 + B] \psi^{(+)}
= k^2  \psi^{(+)}\,,
\label{eqn:iiseventeen}
\end{equation}
for $k \rightarrow 0$. This can be solved explicitly in the exterior region $r > a$ with overall normalization fixed by (\ref{eqn:iiseven}).
We can approximate the interior solution by the exact zero-energy solution of
(\ref{eqn:iiseventeen}) because $k^2$ is a regular 
perturbation of $D^2 + B$ for $r < a$.
Then an interior solution of (\ref{eqn:iiseventeen}) can be expanded as a power
series in $k^2$. Following this the interior and exterior solutions are
matched at 
$r = a$~\cite{x}.
\renewcommand{\thesection}{\Roman{section}}
\renewcommand{\theequation}{\thesection.\arabic{equation}}
\section{LOW ENERGY SCATTERING STATES}
\setcounter{equation}{0}
\renewcommand{\thesection}{\arabic{section}}
\renewcommand{\theequation}{\thesection.\arabic{equation}}

Since the case of noncentral potentials may be unfamiliar we will parallel our discussion with the special case of radial symmetry in the interest of clarity.
\subsection{{\bf Central field: ${\bf r > a}$}}
Expand $\psi^{(+)}({\bf k, r})$ in partial waves,
\begin{equation}
\psi^{(+)}({\bf k},{\bf r})
=
\frac{1}{\sqrt 2\, \pi} \sum_{l=-\infty}^{\infty}\,
\psi_{\scriptscriptstyle{l}}(k, r)
e^{i l (\theta - \Theta)}\,.
\label{eqn:iiione}
\end{equation}
Equations (\ref{eqn:iiione}) and (\ref{eqn:iisix}) give the normalization condition
\begin{equation}
\int_0^{\infty} \,dr r \,
\psi_{\scriptscriptstyle{l}}^{\ast}(k,r)\,
\psi_{\scriptscriptstyle{l}}(k^{\prime},r)
=
\delta(E - E^{\prime})\,.
\label{eqn:iiitwo}
\end{equation}
Substitution of (\ref{eqn:iiione}) and (\ref{eqn:iiten}) in 
(\ref{eqn:iiseventeen}) results in Bessel's equation for  $\psi_{\scriptscriptstyle{l}}$, with $l$ shifted to $l - \Phi/2 \pi$ for $r > a$.
In order to include the case when $\Phi/2 \pi$ is an integer we
choose as linearly independent solutions the Hankel functions
$H_{|l - \Phi/2 \pi|}^{(\pm)}(kr)$
whose asymptotic behavior for $r \rightarrow \infty$ is
\begin{equation}
H_{\nu}^{(\pm)}(kr)
\sim
\sqrt{\frac{2}{\pi k r}}\,
e^{\pm i (k r - \nu \pi/2 - \pi/4)}
\,.
\label{eqn:iiithree}
\end{equation}
Setting $W= |l - \Phi/2 \pi|$ we construct $\psi_{\scriptscriptstyle{l}}$ as the following linear
combination
\begin{equation}
\psi_{\scriptscriptstyle{l}}(k,r)
=
\frac{ e^{- i \pi W/2}\, e^{i \pi |l|} }{2 \sqrt 2} \,
\left(
H_{\scriptscriptstyle{W}}^{(-)}(kr) +
e^{ i \pi (W - |l|)}\, e^{2 i \delta_l(k)}\,
H_{\scriptscriptstyle{W}}^{(+)}(kr)
\right)\,,
\label{eqn:iiifour}
\end{equation}
where $S_l = e^{2 i \delta_l}$ is the $S$-matrix for the partial phase
shift $\delta_l$. Recalling that in two dimensions
\begin{equation}
e^{i {\bf k \cdot r}}
=
e^{i k r \cos(\theta - \Theta)}
=
\sum_{l=-\infty}^{\infty}i^l\,J_l(kr)\,e^{i l (\theta - \Theta)}\,,
\label{eqn:iiifive}
\end{equation}
and noting (\ref{eqn:iiithree}) we see that the normalization factors in 
(\ref{eqn:iiifour}) ensure that (\ref{eqn:iiione}) assumes the asymptotic form
(\ref{eqn:iiseven}) as  $r \rightarrow \infty$.
\subsection{{\bf General field: ${\bf r > a}$}}
Although radial symmetry is present for $r > a$, the absence of rotational 
symmetry for $r < a$ can cause the incident particle to scatter into a final
state that is a superposition of angular momentum states.
Thus the $S$-matrix is no longer diagonal in $l:\, S_l\rightarrow S_{\scriptscriptstyle{l,L}}$,
where $L$ is the initial-state angular momentum. Then  (\ref{eqn:iiione})
generalizes to
\begin{equation}
\psi^{(+)}({\bf k},{\bf r})
=
\frac{1}{\sqrt 2\, \pi} \sum_{l,L}
\,
\psi_{\scriptscriptstyle{l,L}}(k,r) \, e^{i l \theta} \, e^{- i L \Theta}\,,
\label{eqn:iiisix}
\end{equation}
which, together with (\ref{eqn:iisix}), results in the normalization condition
\begin{equation}
\sum_{l}
\int_0^{\infty} \,dr r \,
\psi_{\scriptscriptstyle{l,L}}^{\ast}(k, r)\,
\psi_{\scriptscriptstyle{l,L^{\prime}}}(k^{\prime}, r)
=
\delta_{\scriptscriptstyle{L,L^{\prime}}}\,
\delta(E - E^{\prime})\,.
\label{eqn:iiiseven}
\end{equation}
Equation (\ref{eqn:iiifour}) now generalizes to
\begin{eqnarray}
\psi_{\scriptscriptstyle{l,L}}(k,r) 
&
=
&
e^{- i \pi (W_l + W_L)/4}\, e^{i \pi (|l| + |L|)/2}
\nonumber \\
& &
\times
\frac{1}{2 \sqrt 2}
\left(
\delta_{\scriptscriptstyle{l,L}}\, H_{\scriptscriptstyle{W_l}}^{(-)}(kr)+
e^{i \pi (W_l- |l|)/2}\,S_{\scriptscriptstyle{l,L}}\,e^{i \pi (W_L- |L|)/2}\,
H_{\scriptscriptstyle{W_l}}^{(+)}(kr)
\right)\,,
\label{eqn:iiieight}
\end{eqnarray}
where $W_l= |l - \Phi/2 \pi|$, etc.. Unless $\Theta$ needs to be displayed,
as in (\ref{eqn:iiisix}), we suppress it in what follows. Again, the 
normalization factors in  (\ref{eqn:iiieight}) are chosen so that 
(\ref{eqn:iiisix}) assumes the asymptotic form (\ref{eqn:iiseven}).
The scattering amplitude is given by
\begin{equation}
f({\bf k^{\prime}},{\bf k})
=
\frac{1}{\sqrt{2 \pi k}}
\sum_{l,L}\,(S_{\scriptscriptstyle{l,L}} - 
\delta_{\scriptscriptstyle{l,L}})\,
 e^{i l \theta} \, e^{- i L \Theta}\,
e^{i \pi ( W_L - W_l - 1)/4}\,.
\label{eqn:iiinine}
\end{equation}
\subsection{{\bf General field: ${\bf r < a}$}}
We seek zero-energy solutions of (\ref{eqn:iiseventeen}) in the region $r < a$
that are sufficiently regular to maintain the Hermiticity of $D^2 + B$.
This operator factorizes to $L^{\dagger}L$ so that  (\ref{eqn:iiseventeen})
at $k^2=0$ reduces to
\begin{equation}
L^{\dagger}L \psi = 0\,,
\label{eqn:iiiten}
\end{equation}
where
\begin{equation}
L= 
e^{-i \theta} \, 
\left(
\frac{1}{i}\, \frac{\partial}{\partial r}
- \frac{1}{r}\, \frac{\partial}{\partial \theta}
- \frac{1}{r}\, \frac{\partial \phi}{\partial \theta}
- i \frac{\partial \phi}{\partial r}
\right)\,.
\label{eqn:iiieleven}
\end{equation}
One set of solutions is given by
\begin{equation}
L \psi = 0\,,
\label{eqn:iiitwelve}
\end{equation}
whose solution by inspection is
\begin{equation}
\psi = e^{-\phi(r,\theta)}\,g(re^{-i \theta})\,,
\label{eqn:iiithirteen}
\end{equation}
where $g$ is analytic in $r e^{-i \theta}$ in and on the disk $D$.
Solutions of the form (\ref{eqn:iiithirteen}) do not give all of the regular 
solutions of (\ref{eqn:iiiten}). This is evident in the limit of radial 
symmetry, for then $\psi$ is a superposition of only negative or zero angular 
momentum states.
\paragraph*{}
There are irregular solutions of (\ref{eqn:iiitwelve}) and hence 
(\ref{eqn:iiiten}) of the form
\begin{equation}
\psi
=
e^{-\phi(r,\theta)}\,
h(r^{-1}e^{i \theta})\,,
\label{eqn:iiifourteen}
\end{equation}
where $h$ may be expanded in a power series away from the origin. These 
solutions can be used to find additional regular solutions of 
(\ref{eqn:iiiten}) that reduce to superpositions of positive angular momenta in the radial symmetry limit. Thus, we look for regular solutions about the origin of the form
\begin{equation}
\psi
=
e^{-\phi}\,
h(r^{-1}e^{i \theta})\,
F(r,\theta)\,.
\label{eqn:iiifivteen}
\end{equation}
Then (\ref{eqn:iiiten}) gives
\begin{equation}
L^{\dagger}
\left[
e^{-\phi}\,e^{-i\theta}\,  h(r^{-1}e^{i \theta})\,
\left(
\frac{\partial F}{\partial r}
-\frac{i}{r}\, \frac{\partial F}{\partial \theta}
\right)
\right]
= 0
\,.
\label{eqn:iiisixteen}
\end{equation}
Again by inspection the solution of (\ref{eqn:iiisixteen}) is
\begin{equation}
e^{-\phi}\,e^{-i\theta}\, h\,
\left(
\frac{\partial F}{\partial r}
-\frac{i}{r}\, \frac{\partial F}{\partial \theta}
\right)
= 
e^{\phi}\, b(re^{i \theta})
\,,
\label{eqn:iiiseventeen}
\end{equation}
where $b$ is analytic in $re^{i \theta}$ in and on $D$. Actually, $h$ is now an unnecessary complication.
Letting $F=f(r, \theta)/h$, we get
\begin{equation}
\frac{\partial f}{\partial r}
-\frac{i}{r}\, \frac{\partial f}{\partial \theta}
= 
e^{i \theta}\, e^{2\phi}\, b(re^{i \theta})
\,,
\label{eqn:iiieighteen}
\end{equation}
and hence (\ref{eqn:iiifivteen}) becomes
\begin{equation}
\psi
=
e^{-\phi}\, f \,.
\label{eqn:iiinineteen}
\end{equation}
Equation (\ref{eqn:iiieighteen}) indicates that $f$ is undetermined up to a
function of the form $p (r e^{-i \theta})$. But this is the same as $g$ in 
(\ref{eqn:iiithirteen}), and so we set $p=0$. Also, the value of $\psi$ at the origin can be fixed by $g$. So for definiteness we require $f(0)=0$. Noting 
that 
\begin{equation}
\nabla^2
=
e^{-i \theta}\,
\left(
\frac{\partial}{\partial r}
-\frac{i}{r}\, \frac{\partial}{\partial \theta}
\right)\,
e^{i \theta}\,
\left(
\frac{1}{r}
+\frac{i}{r}\, \frac{\partial}{\partial \theta}
\right)\,,
\label{eqn:iiitwenty}
\end{equation}
the solution of (\ref{eqn:iiieighteen}) is, for ${\bf r} \in D$,
\begin{equation}
f({\bf r})
=
\frac{1}{2 \pi}\,
e^{i \theta}\,
\left(
\frac{\partial}{\partial r}
+\frac{i}{r}\, \frac{\partial}{\partial \theta}
\right)\,
\int_{D}\,d^2 r^{\prime}\, \ln{|{\bf r - r^{\prime}}|}\,
e^{2 \phi({\bf r^{\prime}})}\,
b(r^{\prime}e^{i \theta^{\prime}})
+ C\,,
\label{eqn:iiitwentyone}
\end{equation}
where $C$ is a constant fixed by $f(0)=0$. Since 
\begin{equation}
e^{i \theta}\,
\left(
\frac{\partial}{\partial r}
+\frac{i}{r}\, \frac{\partial}{\partial \theta}
\right)\,
\ln{|{\bf r - r^{\prime}}|}
=
- e^{i \theta^{\prime}}\,
\left(
\frac{\partial}{\partial r^{\prime}}
+\frac{i}{r^{\prime}}\, \frac{\partial}{\partial \theta^{\prime}}
\right)\,
\ln{|{\bf r - r^{\prime}}|}\,,
\label{eqn:iiitwentytwo}
\end{equation}
$f$ takes the final form
\begin{equation}
f({\bf r})
=
- \frac{1}{2 \pi}\,
\int_D\, d^2 r^{\prime} \,e^{i \theta^{\prime}}\,
e^{2 \phi({\bf r^{\prime}})}\,
b(r^{\prime}e^{i \theta^{\prime}})\,
\left(
\frac{\partial}{\partial r^{\prime}}
+\frac{i}{r^{\prime}}\, \frac{\partial}{\partial \theta^{\prime}}
\right)\,
\ln{(|{\bf r - r^{\prime}}|/r^{\prime})}\,.
\label{eqn:iiitwentythree}
\end{equation}
Combining (\ref{eqn:iiithirteen}), (\ref{eqn:iiinineteen}) and  
(\ref{eqn:iiitwentythree}), the general solution of (\ref{eqn:iiiten}) is
\begin{equation}
\psi(r, \theta)
=
e^{-\phi(r,\theta )}\,
(
g(r e^{-i \theta})
+
f(r,\theta )
)\,.
\label{eqn:iiitwentyfour}
\end{equation}
The functions $b$, $f$ and $g$ will be determined below when we join the region $r < a$ with $r > a$. They then acquire energy-dependent normalization factors that depend on scattering data, including the initial-state angular momentum
$L$. Thus $\psi$ in (\ref{eqn:iiitwentyfour}) has an implicit dependence on
$\Theta$. 
\paragraph*{}
As discussed at the end of Sec. II, $k^2$ is a regular perturbation
of $D^2 + B$ in (\ref{eqn:iiseventeen}) in the region $r < a$.
Hence the radial wave functions $\psi_{\scriptscriptstyle{l,L}}(k,r)$ in (\ref{eqn:iiisix}) may be
expanded in $k^2$ for  $r < a$. Inserting the implicit $\Theta$-dependence of
 $\psi(r,\theta)$ in (\ref{eqn:iiitwentyfour}) we expand it in partial waves
\begin{equation}
\psi(r,\theta,\Theta)
=
\sum_{l,L}\,\psi_{\scriptscriptstyle{l,L}}(r)
\,e^{i l \theta}\,e^{-i L \Theta}\,.
\label{eqn:iiitwentyfive}
\end{equation}
We set
\begin{equation}
\psi_{\scriptscriptstyle{l,L}}(k,r)/ \sqrt 2 \pi
=
\psi_{\scriptscriptstyle{l,L}}(r)\,
\left(
1 + k^2 \chi_{\scriptscriptstyle{l,L}}(r) + \mbox{O} (k^4)
\right)
\label{eqn:iiitwentysix}
\end{equation}
and thus 
\begin{equation}
\psi^{(+)}({\bf k,r})
=
\psi(r,\theta,\Theta) +
\frac{k^2}{(2 \pi)^2}\,
\int_0^{2 \pi} d\theta^{\prime} \,
\int_0^{2 \pi}d\Theta^{\prime} \,
\psi(r,\theta-\theta^{\prime},\Theta-\Theta^{\prime})\,
\chi(r,\theta^{\prime},\Theta^{\prime})\, +
\mbox{O} (k^4 \psi)\,,
\label{eqn:iiitwentyseven}
\end{equation}
where $\chi$ can be expanded as in (\ref{eqn:iiitwentyfive}). We will 
abbreviate (\ref{eqn:iiitwentyseven}) as
\begin{equation}
\psi^{(+)}({\bf k,r})
=
\psi(r,\theta) +
k^2 \psi \star \chi + 
\mbox{O} (k^4 \psi)\,,
\label{eqn:iiitwentyeight}
\end{equation}
where the star denotes convolution. An equation for $\chi$ can be obtained
by substituting (\ref{eqn:iiitwentyeight}) in (\ref{eqn:iiseventeen}) and
retaining terms of order $k^2$. As we will see, $\chi$ is not required in the
general field case.
\paragraph*{}
To fix $b$ and $f$ define the operator
\begin{equation}
{\mathcal L}=
\frac{\partial}{\partial r}
-\frac{i}{r}\, \frac{\partial}{\partial \theta}\,,
\label{eqn:iiitwentynine}
\end{equation}
and let it act on $\psi^{(+)}$ in (\ref{eqn:iiitwentyeight}), using 
(\ref{eqn:iiitwentyfour}):
\begin{equation}
{\mathcal L}\,\psi^{(+)}
=
-\psi\,{\mathcal L} \,\phi +
e^{-\phi}\,{\mathcal L} \, f +
k^2 {\mathcal L} \,\psi\, \star \chi +
\mbox{O} (k^4  {\mathcal L} \psi)\,.
\label{eqn:iiithirty}
\end{equation}
Equation (\ref{eqn:iisixteen}) with $R=a$ gives
\begin{equation}
\partial_{\scriptscriptstyle{\theta}}\, 
\phi(a, \theta) = 0\,, \,\,\, (\partial_r \phi(r, \theta))_a =
-\Phi/2 \pi a+ \mbox{O}(||B||R^2/a^2)\,.
\label{eqn:iiithirtyone}
\end{equation} 
From now on the arbitrarily small correction to the radial derivative of
$\phi$ at $r = a$ will be implicit in what follows.
Thus (\ref{eqn:iiithirtyone}) and (\ref{eqn:iiieighteen}) applied to 
(\ref{eqn:iiithirty}) at $r=a$ give
\begin{equation}
{\mathcal L}\,\psi^{(+)}(k,a,\theta)
=
\Phi\, \psi(a,\theta)/2 \pi a + e^{i \theta}\,b(ae^{i \theta}) +
k^2 {\mathcal L} \,\psi\, \star \chi +
\mbox{O} (k^4  {\mathcal L} \psi)\,.
\label{eqn:iiithirtytwo}
\end{equation}
Denote the wave functions on either side of $a$ by $\psi_<^{(+)}$ and
$\psi_>^{(+)}$. Continuity of  $\psi^{(+)}$ and ${\mathcal L}\,\psi^{(+)}$ at
$r=a$ and repeated use of (\ref{eqn:iiitwentyeight}) allow
(\ref{eqn:iiithirtytwo}) to be put in the form
\begin{equation}
e^{i \theta}\,b(ae^{i \theta})
=
({\cal L} - \Phi/2 \pi a)\,\psi_>^{(+)} -
k^2 \Phi\,\psi_>^{(+)}\star\chi/2 \pi a +
k^2 {\mathcal L}\,\psi_>^{(+)}\star\chi +
\mbox{O} (k^4  {\mathcal L}\psi_>^{(+)} )\,.
\label{eqn:iiithirtythree}
\end{equation}
Since $b(re^{i \theta})$ is analytic in and on $D$ we can make the expansion
\begin{equation}
b(re^{i \theta})
=
\sum_{l=0}^{\infty}\,
\sum_{L=-\infty}^{\infty}
b_{\scriptscriptstyle{l,L}} \,r^l \,e^{i l\theta}\,e^{-i L\Theta}\,,
\label{eqn:iiithirtyfour}
\end{equation}
where we have anticipated the $\Theta$-dependence of $b$.
From (\ref{eqn:iiisix}), (\ref{eqn:iiithirtythree}) and 
(\ref{eqn:iiithirtyfour})
 we get
\begin{equation}
\sqrt 2 \pi b_{\scriptscriptstyle{l-1,L}} \,a^{l-1} 
=
\left(
\frac{d}{dr} + (l - \Phi/2 \pi)/r
\right)\,
\psi_{\scriptscriptstyle{l,L}}^>(k,r)\,
( 1 + k^2 \chi_{\scriptscriptstyle{l,L}}(r) + \mbox{O} (k^4) )\,,
\label{eqn:iiithirtyfive}
\end{equation}
with $r=a$ after differentiating.
Referring to (\ref{eqn:iiieight}) it is evident from (\ref{eqn:iiithirtyfive})
that the expansion coefficients $b_{\scriptscriptstyle{l,L}}$ and hence $b$ and $f$ in 
(\ref{eqn:iiitwentythree}) will be determined to leading order in $k^2$ once
$S_{\scriptscriptstyle{l,L}}$ is known.
\paragraph*{}
There now remains the function $g$ in (\ref{eqn:iiitwentyfour}).
Equations (\ref{eqn:iiitwentyfour}) and 
(\ref{eqn:iiitwentyeight}) together with continuity of $\psi^{(+)}$ at $r=a$
give
\begin{equation}
\psi_>^{(+)}(k,a,\theta)=
g(a e^{-i \theta}) + f(a,\theta) +
k^2 \psi\star \chi(a,\theta) + \mbox{O} (k^4 \psi)\,.
\label{eqn:iiithirtysix}
\end{equation}
Letting 
\begin{equation}
f(r,\theta)=
\sum_{l,L} f_{\scriptscriptstyle{l,L}}(r)\,e^{i l \theta}\,e^{-i L \Theta}\,,
\label{eqn:iiithirtyseven}
\end{equation}
and recalling that $g$ is analytic in and on $D$ so that
\begin{equation}
g(r e^{-i \theta})=
\sum_{l=0}^{\infty}\sum_{L=-\infty}^{\infty}
g_{\scriptscriptstyle{l,L}}\,r^l \,e^{-i l \theta}\,e^{-i L \Theta}\,,
\label{eqn:iiithirtyeight}
\end{equation}
we obtain from (\ref{eqn:iiisix}),  
(\ref{eqn:iiithirtysix})-(\ref{eqn:iiithirtyeight}), for $l\ge 0$,
\begin{equation}
\psi_{\scriptscriptstyle{-l,L}}^{>}(k,a)/\sqrt2 \pi
=
g_{\scriptscriptstyle{l,L}}\,a^l
+ f_{\scriptscriptstyle{-l,L}}(a) +
k^2 \psi_{\scriptscriptstyle{-l,L}}^{>}(k,a)\,\chi_{\scriptscriptstyle{-l,L}}(a)/\sqrt2 \pi + 
\mbox{O} (k^4 \psi_{\scriptscriptstyle{-l,L}}^{>})\,.
\label{eqn:iiithirtynine}
\end{equation}
This simplifies on making the expansion
\begin{equation}
\ln(r^2 + r^{\prime 2} - 2 r  r^{\prime} \cos(\theta - \theta^{\prime}))=
\ln r_>^2 - 
2 \sum_{l=1}^{\infty}\frac{1}{l}\,\left(\frac{r_<}{r_>}\right)^l\,
\cos[l(\theta - \theta^{\prime})]\,,
\label{eqn:iiiforty}
\end{equation}
in (\ref{eqn:iiitwentythree}), giving $f_{\scriptscriptstyle{-l,L}}(a)=0$, $l>0$ and
\begin{equation}
f_{\scriptscriptstyle{0,L}}(a)
=
\frac{1}
{2 \pi}
\int_0^a dr
\int_0^{2 \pi} d\theta\, e^{i \theta}\,e^{2 \phi(r,\theta)}\,
\sum_{l=0}^{\infty}b_{\scriptscriptstyle{l,L}}\,r^l \,e^{i l \theta}\,.
\label{eqn:iiifortyone}
\end{equation}
Thus $g$ is determined to leading order in  $k^2$ by 
(\ref{eqn:iiithirtyeight})-(\ref{eqn:iiithirtynine}) once 
$S_{\scriptscriptstyle{l,L}}$ is known.
We have now fully determined the low energy limit of 
$\psi^{(+)}({\bf k, r})$ for general magnetic fields in terms of the 
$S$-matrix.
\subsection{{\bf Central Field: ${\bf r < a}$}}
Now everything is diagonal, and we may set $a=R$. 
Refer back to (\ref{eqn:iiifour}) and define
the energy-dependent part, $\Delta_{\scriptscriptstyle{l}}$, of the phase shifts by
\begin{equation}
\delta_{\scriptscriptstyle{l}}(k^2)
=
\pi (|l| - |l - \Phi/2 \pi|)/2 + \Delta_l(k^2) + m \pi\,,
\label{eqn:iiifortytwo}
\end{equation}
where $m = 0, \pm 1, ..$. Then (\ref{eqn:iiifour}) reduces to, for 
$r > a$,
\begin{equation}
\psi_l(k,r)
=
2^{-1/2}\,(-1)^m\,i^{|l|}\, e^{i \delta_l}\,
(J_{\scriptscriptstyle{W}}(kr)\,\cos \Delta_{\scriptscriptstyle{l}}
- Y_{\scriptscriptstyle{W}}(kr) \,\sin \Delta_{\scriptscriptstyle{l}})\,,
\label{eqn:iiifortythree}
\end{equation}
where $Y_W$ is the Bessel function of the second kind. Then at
$r=a$ with $ka \ll 1$ and $W \neq 0$,
\begin{eqnarray}
\psi_l^>(k,a)
&=&
2^{-1/2}\,(-1)^m\,i^{|l|}\, e^{i \delta_l}\,
[
(ka/2)^W/\Gamma(W+1) + \Delta_{\scriptscriptstyle{l}} 
\,\Gamma(W)\,(ka/2)^{-W}/\pi
] \nonumber \\
& & \times( 1 +  \mbox{O} (k^2,\Delta_l^2) )\,,
\label{eqn:iiifortyfour}
\end{eqnarray}
\begin{eqnarray}
(r \partial_r \,\psi_l^>)_a
&=&
2^{-1/2}\,W\,(-1)^m\,i^{|l|}\, e^{i \delta_{\scriptscriptstyle{l}}}\,
[
(ka/2)^W/\Gamma(W+1) -\Delta_{\scriptscriptstyle{l}} 
\,\Gamma(W)\,(ka/2)^{-W}/\pi
] \nonumber \\
& &\times ( 1 +  \mbox{O} (k^2,\Delta_{\scriptscriptstyle{l}}^2) )\,.
\label{eqn:iiifortyfive}
\end{eqnarray}
It will be shown in Sec. IV that $\Delta_{\scriptscriptstyle{l}}= \mbox{O}(ka)^{2W}$ at least.
In (\ref{eqn:iiifortyfour}) and (\ref{eqn:iiifortyfive}) the remainder term
$\mbox{O} (k^2,\Delta_{\scriptscriptstyle{l}}^2)$ should be replaced with 
$\mbox{O} ((ka)^2\, \ln ka)$ when $W=1$.
\paragraph*{}
We can now calculate $b_{\scriptscriptstyle{l-1}}$ in (\ref{eqn:iiithirtyfive}).
For $l > \Phi/2 \pi$, (\ref{eqn:iiithirtyfive}), (\ref{eqn:iiifortyfour}) and
(\ref{eqn:iiifortyfive}) give
\begin{equation}
b_{\scriptscriptstyle{l-1}}\,a^l
=
\pi^{-1}\, (l - \Phi/2 \pi)\,(-1)^m\,i^{l}\, e^{i \delta_l}\,
(ka/2)^W/\Gamma(W+1)\, ( 1 +  \mbox{O} (k^2,\Delta_l^2) )\,,
\label{eqn:iiifortysix}
\end{equation}
and for $ 1 \le l < \Phi/2 \pi$,
\begin{equation}
b_{\scriptscriptstyle{l-1}}\,a^l
=
\pi^{-2}\, (l - \Phi/2 \pi)\,(-1)^m\,i^{l}\, 
e^{i \delta_{\scriptscriptstyle{l}}}\,\Delta_{\scriptscriptstyle{l}}\,
\Gamma(W)
(ka/2)^{-W}\, ( 1 +  \mbox{O} (k^2,\Delta_{\scriptscriptstyle{l}}^2) )\,.
\label{eqn:iiifortyseven}
\end{equation}
For the case $W=0$,
\begin{equation}
b_{\scriptscriptstyle{l-1}}\,a^l
=
-\pi^{-2}\,(-1)^m\,i^{l}\, e^{i \delta_{\scriptscriptstyle{l}}}\,
\Delta_{\scriptscriptstyle{l}}\,
( 1 +  \mbox{O} (\Delta_{\scriptscriptstyle{l}}^2) )\,.
\label{eqn:iiifortyeight}
\end{equation}
For $g_l$, (\ref{eqn:iiithirtynine}) gives
\begin{equation}
g_l
=
\frac{ \psi_{\scriptscriptstyle{-l}}^>(k,a)}{\sqrt 2 \pi a^l}\,
(
1 - k^2 \chi_{\scriptscriptstyle{-l}}(a) +  \mbox{O} (k^4)
)\,,
\label{eqn:iiifortynine}
\end{equation}
since (\ref{eqn:iiifortyone}) gives $f_{\scriptscriptstyle{0,L}}(a)=0$ for the case of radial 
symmetry.
Combining (\ref{eqn:iiifortynine}) with (\ref{eqn:iiifortyfour}), we obtain,
for $l \ge 0$,
\begin{eqnarray}
g_l
&
=
& 
(-1)^m\,i^{l}\, e^{i \delta_{\scriptscriptstyle{-l}}} \,(2 \pi a^l)^{-1}\,
[
(ka/2)^{l + \Phi/ 2 \pi}/\Gamma(l + 1+ \Phi/ 2 \pi)
\nonumber \\
& &- \pi^{-1}\,\Delta_{\scriptscriptstyle{-l}}\,\Gamma(l + \Phi/ 2 \pi)\,
(ka/2)^{-(l + \Phi/ 2 \pi)}]
\times
( 1 +  \mbox{O} (k^2,\Delta_{\scriptscriptstyle{-l}}^2) )\,.
\label{eqn:iiifivty}
\end{eqnarray}
\paragraph*{}
Referring to  (\ref{eqn:iiithirtyseven}), (\ref{eqn:iiiforty}) and
  (\ref{eqn:iiitwentythree}) one finds for $r \le a$,
\begin{eqnarray}
f_l(r) & = &
b_{\scriptscriptstyle{l-1}}\,r^{-l} \,
\int_0^r dx \, x^{2l -1}\, e^{2 \phi (x)},\;\; l \ge 1 \nonumber \\
 & =& 0,\;\; l \le 0\,.
\label{eqn:iiifivtyone}
\end{eqnarray}
In the radial symmetry limit $\psi_{\scriptscriptstyle{l,L}}(r)$ in (\ref{eqn:iiitwentysix})
 becomes diagonal, with (\ref{eqn:iiitwentyfour}) now giving, for $r \le a$,
\begin{eqnarray}
\psi_l(r) & = &
b_{\scriptscriptstyle{l-1}}\,r^{-l} \,e^{- \phi(r)}\,
\int_0^r dx \, x^{2l -1}\, e^{2 \phi (x)} ,\; l \ge 1 \nonumber \\
& = & g_{\scriptscriptstyle{-l}}\,r^{-l} \,e^{-\phi(r)},\; l \le 0\,.
\label{eqn:iiifivtytwo}
\end{eqnarray}
Since it will be needed in what follows we end this section by calculating 
$\chi_l$ in (\ref{eqn:iiitwentysix}). Substitution of 
 (\ref{eqn:iiitwentyfive}) and (\ref{eqn:iiitwentysix}) in 
(\ref{eqn:iiseventeen}) and matching terms of $ \mbox{O} (k^2)$ gives
\begin{equation}
\left(
-\frac{d}{dr} + \frac{l-1}{r} + \phi^{\prime}
\right)
\left(
\frac{d}{dr} + \frac{l}{r} + \phi^{\prime}
\right)\,
\psi_l\,\chi_l = \psi_l\,.
\label{eqn:iiifivtythree}
\end{equation}
Requiring $\chi_l(0)=0$, (\ref{eqn:iiifivtytwo}) and  (\ref{eqn:iiifivtythree})
fix $\chi_l$ for $ r \le a$ to be
\begin{eqnarray}
\chi_l & = &
- \int_0^r dx x^{-2 |l| -1}\,e^{2 \phi(x)}
\int_0^x dy\,y^{2 |l| +1}\,e^{-2 \phi(y)}\,,\; l \le 0 \nonumber \\
&= &
- \int_0^r dx x^{2 l -1}\,e^{2 \phi(x)}
\left(
\int_0^x dw\,w^{2 l -1}\,e^{2 \phi(w)}
\right)^{-2}\nonumber \\
& &\times
\int_0^x dy\,y^{1- 2 l}\,e^{2 \phi(y)} 
\left(
\int_0^y dz\,z^{2 l-1}\,e^{2 \phi(z)}
\right)^{2}\,,\; l \ge 1\,.
\label{eqn:iiifivtyfour}
\end{eqnarray}
It is important to bound the growth of $\chi_l$ for $|l| \rightarrow \infty$.
In the limit of radial symmetry (\ref{eqn:iisixteen}) reduces to
\begin{equation}
\phi(r)
=
-\int_0^a
d r^{\prime}\, r^{\prime}\, B(r^{\prime})\,\ln(r_>/a)\,,
\label{eqn:iiifivtyfive}
\end{equation}
for $r \le a$.
Then some easy estimates applied to (\ref{eqn:iiifivtyfour}) and 
(\ref{eqn:iiifivtyfive}) yield
\begin{equation}
|\phi(r)| \le ||B||\,(a - r)/2 \sqrt{\pi}\,,
\label{eqn:iiifivtysix}
\end{equation}
and
\begin{eqnarray}
|\chi(r)| & \leq & 
\frac{e^{2 ||B|| a \sqrt{\pi}}}{4 (|l| + 1)}\,r^2\,,\;\; l 
\leq 0 \nonumber \\
& \leq &
\frac{e^{6||B|| a \sqrt{\pi}}}{4 (l + 1)}\,r^2\,,\;\; l \geq 1\,.
\label{eqn:iiifivtyseven}
\end{eqnarray}
\renewcommand{\thesection}{\Roman{section}}
\renewcommand{\theequation}{\thesection.\arabic{equation}}
\section{{\bf LOW ENERGY PHASE SHIFTS}}
\setcounter{equation}{0}
\renewcommand{\thesection}{\arabic{section}}
\renewcommand{\theequation}{\thesection.\arabic{equation}}

In order to calculate (\ref{eqn:iitwelve}) in the limit $m^2 \rightarrow 0$ the
leading energy-dependent behavior of $S_{\scriptscriptstyle{l,L}}$ is required.
The case of central fields is dealt with first.
\subsection{{\bf Central Fields}}
The calculation of $\Delta_{\scriptscriptstyle{l}}$ in (\ref{eqn:iiifortytwo}) proceeds by matching the log-derivatives 
$\gamma_{\scriptscriptstyle{l}}=r \partial_r \ln \psi_{\scriptscriptstyle{l}}$ at $r=a$, where again we can set $a=R$..
Then (\ref{eqn:iiifortythree}) gives
\begin{equation}
\tan \Delta_{\scriptscriptstyle{l}}
=
\frac{ 
\gamma_{\scriptscriptstyle{l}} J_{\scriptscriptstyle{W}}(ka) -
ka J_{\scriptscriptstyle{W}}^{\prime}(ka)
}
{
\gamma_{\scriptscriptstyle{l}} Y_{\scriptscriptstyle{W}}(ka) -
ka Y_{\scriptscriptstyle{W}}^{\prime}(ka)
}\,,
\label{eqn:ivone}
\end{equation}
where  $\gamma_{\scriptscriptstyle{l}}$ denotes  $\gamma_{\scriptscriptstyle{l}}$ (inside).
For $ka \ll 1$ this reduces to
\begin{equation}
\Delta_{\scriptscriptstyle{l}}
=
\pi\, \frac{ W - \gamma_{\scriptscriptstyle{l}} }
{ W +\gamma_{\scriptscriptstyle{l}} }\,
\frac{ (ka/2)^{2W} }{\Gamma(W) \Gamma(W+1)}\,
(1 + \mbox{O}(ka)^2)\,,
\label{eqn:ivtwo}
\end{equation}
for $W=|l - \Phi/2 \pi| \neq 0, 1, ..$.
Results for integer values of $W$ will be given below.
\paragraph*{}
Now suppose $l \geq 1$. Then (\ref{eqn:iiithirtyfive}) reduces to
\begin{equation}
\gamma_{\scriptscriptstyle{l}}
=
\Phi/2 \pi - l + \sqrt 2 \pi \,b_{\scriptscriptstyle{l-1}}\,a^l / 
\psi_{\scriptscriptstyle{l}}^<(k,a)
+ k^2 [ (\Phi/2 \pi - l - \gamma_{\scriptscriptstyle{l}})\,\chi_{\scriptscriptstyle{l}}(a)
-(r \partial_r \chi_{\scriptscriptstyle{l}})_a ] +\mbox{O}(k^4)\,.
\label{eqn:ivthree}
\end{equation}
From (\ref{eqn:iiisix}), (\ref{eqn:iiitwentyfour}) and 
(\ref{eqn:iiitwentysix}),
\begin{equation}
\psi_{\scriptscriptstyle{l}}^<(k,a)/ \sqrt 2 \pi
=
f_{\scriptscriptstyle{l}}(a) ( 1 + k^2 \chi_{\scriptscriptstyle{l}}(a)  +\mbox{O}(k^4))\,,
\label{eqn:ivfour}
\end{equation}
which, together with  (\ref{eqn:iiifivtyone}) and  (\ref{eqn:ivthree}), gives
\begin{equation}
\gamma_{\scriptscriptstyle{l}}
=
\Phi/2 \pi - l +
\frac{a^{2l}}
{\int_0^a dr\,r^{2l-1}\,e^{2 \phi(r)}}
-k^2\,a \chi_{\scriptscriptstyle{l}}^{\prime}(a) + \mbox{O}(k^4)\,.
\label{eqn:ivfive}
\end{equation}
Note that $\gamma_{\scriptscriptstyle{l}} \sim l$ as $l \rightarrow \infty$, as it 
should.
\paragraph*{}
Next, let $l \leq 0$. Equations (\ref{eqn:iiisix}), (\ref{eqn:iiitwentyfour}) and
(\ref{eqn:iiitwentysix}) give
\begin{equation}
\gamma_{\scriptscriptstyle{l}}
=
\Phi/2 \pi - l +
k^2\,a \chi_{\scriptscriptstyle{l}}^{\prime}(a) + \mbox{O}(k^4)\,,
\label{eqn:ivsix}
\end{equation}
since $f_{\scriptscriptstyle{l}}=0$ for $l \leq 0$ by (\ref{eqn:iiifivtyone}).
Equations (\ref{eqn:iiifortytwo}), (\ref{eqn:ivtwo}), (\ref{eqn:ivfive}) and
(\ref{eqn:ivsix}) determine the leading energy dependence of the phase shifts for $W \neq 0, 1,..$. 
\paragraph*{}
Finally, let $W = 0, 1,..$. There is nothing new in principle here; only
the expansion of $Y_{\scriptscriptstyle{W}}$ for $ka \ll 1$ has to be modified. The whole calculation goes forward as above with the result that for 
$W= 2, 3, \dots, \Delta_{\scriptscriptstyle{l}}$ is still given by (\ref{eqn:ivtwo});
 for $W=1$ replace $\mbox{O}(ka)^2$ in (\ref{eqn:ivtwo}) by 
$\mbox{O}((ka)^2 \ln(ka))$, and for $W=0$,
\begin{equation}
\Delta_{\scriptscriptstyle{l}}
=
\frac{\pi}{2 \ln(ka)} + \mbox{O}(1/\ln^2(ka))\,.
\label{eqn:ivseven}
\end{equation}
It is interesting that the energy dependence of 
$\Delta_{\scriptscriptstyle{l}}$ for
$W=0$ specialized to $l=0$ is exactly the same as that derived by Chandon 
et al.~\cite{xii} for a large class of non-magnetic Schr\"{o}dinger operators 
in 2+1 dimensions.
Note that there is no smooth interpolation of $\Delta_{\scriptscriptstyle{l}}$
from $W \neq 0$ to $W = 0$. This case will, therefore, have to be considered
seperately in what follows.
\subsection{{\bf General Fields}}
The $S$-matrix $S_{\scriptscriptstyle{l,L}}$ appearing in (\ref{eqn:iiieight}) is
obtained from
\begin{equation}
S_{\scriptscriptstyle{l,L}}
=
\delta_{\scriptscriptstyle{l,L}}
+ \sqrt 2 \pi \,i^{-l-1}
\sum_m
\int_0^{\infty}dr\,r\,J_{\scriptscriptstyle{l}}(kr)\, 
V_{\scriptscriptstyle{l-m}}(r)\,
\psi_{\scriptscriptstyle{m,L}}(k,r)\,,
\label{eqn:iveight}
\end{equation}
with $\psi_{\scriptscriptstyle{l,L}}$ given by (\ref{eqn:iiisix}) and 
(\ref{eqn:iieight}) and where
\begin{equation}
V_{\scriptscriptstyle{l-m}}(r)=
\frac{1}{2 \pi}\, 
\int_0^{2 \pi} d\theta\,V(r,\theta)\,e^{-i (l-m) \theta}\,,
\label{eqn:ivnine}
\end{equation}
with $V$ as in (\ref{eqn:iinine}).
An infinite set of coupled equations must be solved to extract the phase shifts in the general field case.
In practice, only a few off-diagonal elements of  $S_{\scriptscriptstyle{l,L}}$
are required to obtain the phase shifts in the low energy limit.
\paragraph*{}
Consider an incident low-energy particle ($ ka \ll 1$) with angular momentum $L$ with respect to the scattering center. It will encounter a high centrifugal barrier $(l - \Phi/2 \pi)^2/r^2$ to the spatially asymmetric region $ r < a$
where $ B(r, \theta) \neq 0$. For values of $L \sim \Phi/2 \pi$ the
barrier is minimized, and so we expect the magnitude of the energy-dependent
corrections to the Aharonov-Bohm phase shifts, $\Delta_{\scriptscriptstyle{L}}$, will
assume their maximum values, as (\ref{eqn:ivtwo}) and (\ref{eqn:ivseven}) 
illustrate
in the centrally symmetric case. The intuition is that  $S_{\scriptscriptstyle{l,L}}$ 
only has significant off-diagonal elements for values of $l, L$ clustered
about $\Phi/2 \pi$ and that otherwise $S_{\scriptscriptstyle{l,L}}$ can be assumed 
diagonal with small error.
\paragraph*{}
To test this hypothesis we will assume that
\begin{equation}
L < \frac{\Phi}{2 \pi} < L+ 1\,,
\label{eqn:ivten}
\end{equation}
and take $S_{\scriptscriptstyle{l,L}}$ to be a $2 \times 2$ matrix to include the 
transitions $L \longleftrightarrow L +1$, and diagonal otherwise.
This $S$-matrix can be calculated for all $\Phi$ satisfying (\ref{eqn:ivten})
 using the results of the previous sections. As one would expect, the mixing of angular momentum states $L$ and $L+1$ is maximum at the mid-interval value 
$\Phi/2 \pi = L + 1/2$. Instead of reproducing this calculation it is more instructive to set $\Phi/2 \pi = L + 1/2$, where the Hankel functions assume a simple form, and show that $\Delta_L$ and $\Delta_{L+1}$ have the same energy 
dependence as in the centrally symmetric case; only the numerical 
coefficients are modified.
In the case of higher order transitions $|\Delta L|  > 1$ involving larger
matrices, we find that the relevant mixing parameters (see below) compared to 
the
 $|\Delta L|  = 1$ case are smaller by factors of order $k^{|\Delta L| - 1}$.
\paragraph*{}
The calculation begins by noting that the potential in (\ref{eqn:iinine}) is
not time-reversal invariant for a fixed magnetic field. Therefore, 
$S_{\scriptscriptstyle{l,L}}$ is not symmetric. We choose the parameterization
\begin{equation}
S=\left(
\begin{array}{cc}
e^{2 i \delta_{L}} \cos 2 \epsilon & i e^{i \alpha} \sin 2 \epsilon \\
i e^{i \beta} \sin 2 \epsilon  & 
e^{2 i \delta_{L+1}} \cos 2 \epsilon
\end{array}
\right) \,,
\label{eqn:iveleven}
\end{equation}
where we expect the mixing parameter $\epsilon$ to vanish as $k \rightarrow 0$.
Unitarity requires $\alpha$ and $\beta$ to be real with
\begin{equation}
\alpha + \beta
=
2(  \delta_{\scriptscriptstyle{L}} + \delta_{\scriptscriptstyle{L+1}} )\,.
\label{eqn:ivtwelve}
\end{equation}
The definition of the phase shifts in (\ref{eqn:iveleven}) is that of Stapp 
et al.~\cite{xiii}, generalized here to include $T$-violation. Referring
to (\ref{eqn:iiifortytwo}) we can rewrite (\ref{eqn:iveleven}) as
\begin{equation}
S
=
\left(
\begin{array}{cc}
e^{2 i (\Delta_{\scriptscriptstyle{L}}-\Phi/4)} \cos 2 \epsilon
&
i e^{ i (\Delta_{\scriptscriptstyle{L}}+ \Delta_{\scriptscriptstyle{L+1}}+
\lambda)}
\sin 2 \epsilon \\
 i e^{ i (\Delta_{\scriptscriptstyle{L}}+ \Delta_{\scriptscriptstyle{L+1}}-
\lambda)}
\sin 2 \epsilon
&
e^{2 i (\Delta_{\scriptscriptstyle{L+1}}+\Phi/4)} \cos 2 \epsilon
\end{array}
\right)\,,
\label{eqn:ivthirteen}
\end{equation}
which introduces a real $T$-violating parameter $\lambda$. From
(\ref{eqn:iiieight}) and  (\ref{eqn:ivtwelve}) with $\Phi/2 \pi = L + 1/2$, the
matching of the interior and exterior log-derivatives at $r=a$ gives
\begin{equation}
\gamma_{\scriptscriptstyle{L,L}}
=
\frac{ ka H_{\scriptscriptstyle{1/2}}^{(-)\,\prime} + e^{2 i\Delta_{\scriptscriptstyle{L}}}
ka H_{\scriptscriptstyle{1/2}}^{(+)\,\prime} \cos 2 \epsilon}
{H_{\scriptscriptstyle{1/2}}^{(-)} + e^{2 i\Delta_{\scriptscriptstyle{L}}}
H_{\scriptscriptstyle{1/2}}^{(+)}  \cos 2 \epsilon} \,,
\label{eqn:ivfourteen}
\end{equation}
with $\gamma_{\scriptscriptstyle{L+1,L+1}} = 
\gamma_{\scriptscriptstyle{L}} 
( \Delta_{\scriptscriptstyle{L}} \rightarrow \Delta_{\scriptscriptstyle{L+1}})$. Also
\begin{equation}
\gamma_{\scriptscriptstyle{L,L+1}} = 
\gamma_{\scriptscriptstyle{L+1,L}} = 
ka  H_{\scriptscriptstyle{1/2}}^{(+)\,\prime} /H_{\scriptscriptstyle{1/2}}^{(+)}\,.
\label{eqn:ivfivteen}
\end{equation}
Recalling that
\begin{equation}
H_{\scriptscriptstyle{1/2}}^{(\pm)}(z)
=
\mp i \,\sqrt{ \frac{2}{\pi z}} \,e^{\pm i z}\,,
\label{eqn:ivsixteen}
\end{equation}
(\ref{eqn:ivthirteen}) becomes,
\begin{equation}
\gamma_{\scriptscriptstyle{L, L}}
=
\frac{
(\frac{1}{2} + i ka)\,e^{-2i ka} + 
(- \frac{1}{2} + i ka)e^{2 i  \Delta_{\scriptscriptstyle{L}}}\cos 2 \epsilon}
{ e^{2 i  \Delta_{\scriptscriptstyle{L}}}\cos 2 \epsilon - e^{-2i ka} }\,,
\label{eqn:ivseventeen}
\end{equation}
\begin{eqnarray}
\gamma_{\scriptscriptstyle{L, L+1}}
&=
-\frac{1}{2}&\;  + \;i z\,.
\label{eqn:iveighteen}
\end{eqnarray}
\paragraph*{}
To get the interior values of $\gamma_{\scriptscriptstyle{l, L}}$, refer to
(\ref{eqn:iiithirtyfive}). Then
\begin{equation}
\gamma_{\scriptscriptstyle{L, L}}
=
\Phi/2 \pi -L +
\frac{ \sqrt 2\,\pi \,a^L\,b_{\scriptscriptstyle{L-1, L}} }
{\psi_{\scriptscriptstyle{L, L}}^<(k,a) }
+ \mbox{O}(ka)^2\,.
\label{eqn:ivnineteen}
\end{equation}
From (\ref{eqn:iiisix}) and (\ref{eqn:iiithirtysix}) - 
(\ref{eqn:iiithirtyeight}),
\begin{equation}
\psi_{\scriptscriptstyle{L, L}}^<(k,a)/\sqrt 2 \pi
=
f_{\scriptscriptstyle{L, L}}(a) + \mbox{O}(ka)^2\,,
\label{eqn:ivtwenty}
\end{equation}
so that
\begin{equation}
\gamma_{\scriptscriptstyle{L, L}}
=
\Phi/2 \pi -L +
\frac{ a^L\,b_{\scriptscriptstyle{L-1, L}} }
{f_{\scriptscriptstyle{L, L}}(a)}\,( 1+ \mbox{O}(ka)^2)
+ \mbox{O}(ka)^2\,.
\label{eqn:ivtwentyone}
\end{equation}
Referring back to (\ref{eqn:iiitwentythree}), (\ref{eqn:iiithirtyseven})
and   (\ref{eqn:iiiforty}) it follows that, for $L > 0$,
\begin{equation}
f_{\scriptscriptstyle{L, L}}(a)
=
\frac{1}{2 \pi \,  a^L}
\int_0^a dr\,r^L
\int_0^{2 \pi}d \theta\,e^{i (1 -L)\theta}\,e^{2 \phi(r,\theta)}
\sum_{m=0}^{\infty} b_{\scriptscriptstyle{m, L}}\,r^m\,e^{i m \theta}\,.
\label{eqn:ivtwentytwo}
\end{equation}
In the sum over $b_{\scriptscriptstyle{m, L}}$ in (\ref{eqn:ivtwentytwo}), only
$b_{\scriptscriptstyle{L-1, L}}$ and $b_{\scriptscriptstyle{L, L}}$ are nonzero as seen from
(\ref{eqn:iiieight}), (\ref{eqn:iiithirtyfive}) and (\ref{eqn:ivthirteen})
since mixing is only assumed for $L \longleftrightarrow L + 1$.
Then (\ref{eqn:ivtwentyone}) and (\ref{eqn:ivtwentytwo}) give
\begin{eqnarray}
\gamma_{\scriptscriptstyle{L, L}}
&
=
\frac{1}{2}
+
2 \pi\, a^{2 L}
&
\left(
\int_0^a dr\,r^{2L -1}
\int_0^{2 \pi}d \theta\,e^{2 \phi(r,\theta)} +
\frac{b_{\scriptscriptstyle{L, L}}}{b_{\scriptscriptstyle{L-1, L}}}\,
\int_0^a dr\,r^{2L}
\int_0^{2 \pi}d \theta\,e^{i\theta}\,e^{2 \phi(r,\theta)}
\right)^{-1} \nonumber \\
& &
\times 
( 1+ \mbox{O}(ka)^2)
+ \mbox{O}(ka)^2\,.
\label{eqn:ivtwentythree}
\end{eqnarray}
Repeating the above steps we find 
\begin{eqnarray}
\gamma_{\scriptscriptstyle{L+1, L+1}}
&
=
- \frac{1}{2}&\;
+\;
2 \pi\, a^{2 L+ 2}
\left(
\int_0^a dr\,r^{2L +1}
\int_0^{2 \pi}d \theta\,e^{2 \phi(r,\theta)} 
\right.\nonumber \\
& &
+\left.\frac{b_{\scriptscriptstyle{L-1, L+1}}}{b_{\scriptscriptstyle{L, L+1}}}\,
\int_0^a dr\,r^{2L}
\int_0^{2 \pi}d \theta\,e^{-i\theta}\,e^{2 \phi(r,\theta)}
\right)^{-1} 
\times 
( 1+ \mbox{O}(ka)^2)
+ \mbox{O}(ka)^2\,.
\label{eqn:ivtwentyfour}
\end{eqnarray}
\paragraph*{}
We now calculate the ratios of $b_{\scriptscriptstyle{l, L}}$ in 
(\ref{eqn:ivtwentythree}) and (\ref{eqn:ivtwentyfour}).
From (\ref{eqn:iiithirtyfive}),
\begin{equation}
\sqrt 2 \pi\,a^l\,b_{\scriptscriptstyle{l-1, L}}
=
(\gamma_{\scriptscriptstyle{l, L}} + L -\Phi/2 \pi)\,
\psi_{\scriptscriptstyle{l, L}}^>(k,a)\,( 1+ \mbox{O}(ka)^2)\,.
\label{eqn:ivtwentyfive}
\end{equation}
Then (\ref{eqn:ivtwentyfive}), (\ref{eqn:ivfivteen}) and 
(\ref{eqn:iveighteen}) give
\begin{equation}
b_{\scriptscriptstyle{L, L}}/b_{\scriptscriptstyle{L-1, L}}
=
\frac{ i k \psi_{\scriptscriptstyle{L+1, L}}^>(k,a)}
{(\gamma_{\scriptscriptstyle{L, L}}- \frac{1}{2})\,\psi_{\scriptscriptstyle{L, L}}^>(k,a)}\,
( 1+ \mbox{O}(ka)^2)\,.
\label{eqn:ivtwentysix}
\end{equation}
Equations (\ref{eqn:iiieight}), (\ref{eqn:ivthirteen}) and (\ref{eqn:ivsixteen})
give
\begin{equation}
\psi_{\scriptscriptstyle{L+1, L}}^>(k,a)/\psi_{\scriptscriptstyle{L, L}}^>(k,a)
=
\frac{
e^{i (\Delta_{\scriptscriptstyle{L}} + \Delta_{\scriptscriptstyle{L+1}} - \lambda - \pi L)}
\sin 2 \epsilon }
{e^{-2i ka} - e^{2 i \Delta_{\scriptscriptstyle{L}}} \cos 2 \epsilon}\,.
\label{eqn:ivtwentyseven}
\end{equation}
Likewise,
\begin{equation}
b_{\scriptscriptstyle{L-1, L+1}}/b_{\scriptscriptstyle{L, L+1}}
=
\frac{ (i ka -1)\,a\,\psi_{\scriptscriptstyle{L, L+1}}^>(k,a)}
{(\gamma_{\scriptscriptstyle{L+1, L+1}} + \frac{1}{2})\,\psi_{\scriptscriptstyle{L+1, L+1}}^>(k,a)}\,
( 1+ \mbox{O}(ka)^2)\,,
\label{eqn:ivtwentyeight}
\end{equation}
\begin{equation}
\psi_{\scriptscriptstyle{L, L+1}}^>(k,a)/\psi_{\scriptscriptstyle{L+1, L+1}}^>(k,a)
=
\frac{
e^{i (\Delta_{\scriptscriptstyle{L}} +\Delta_{\scriptscriptstyle{L+1}} + \lambda - \pi L)}
\sin 2 \epsilon }
{e^{2 i \Delta_{\scriptscriptstyle{L+1}}} \cos 2 \epsilon - e^{-2i ka} }\,.
\label{eqn:ivtwentynine}
\end{equation}
Then solving (\ref{eqn:ivtwentythree}), (\ref{eqn:ivtwentysix}) and  
(\ref{eqn:ivtwentyseven}) for $\gamma_{\scriptscriptstyle{L,L}}$ and matching the 
result with $\gamma_{\scriptscriptstyle{L,L}}$ in (\ref{eqn:ivseventeen}) gives
\begin{eqnarray}
& (1 + i ka)\,e^{-2 i ka}\,I_{\scriptscriptstyle{L}}
+
(i ka -1)\,I_{\scriptscriptstyle{L}}\,e^{2 i\Delta_{\scriptscriptstyle{L}}}\,
 \cos 2 \epsilon =&
\nonumber \\
& 
e^{2 i\Delta_{\scriptscriptstyle{L}}}\,\cos 2 \epsilon -
e^{-2 i ka} +
i k a \,J_{\scriptscriptstyle{L}}\,
e^{i (\Delta_{\scriptscriptstyle{L}} + \Delta_{\scriptscriptstyle{L+1}} - \lambda - \pi L)}\,
\sin 2 \epsilon
+ \mbox{O}((ka)^3, (ka)^2\Delta_{\scriptscriptstyle{L}})\,,&
\label{eqn:ivthirty}
\end{eqnarray}
where
\begin{eqnarray}
I_{\scriptscriptstyle{L}} & = &
(2 \pi\,a^{2L})^{-1}\,
\int_0^a dr\,r^{2L-1}
\int_0^{2 \pi}d \theta\,e^{2 \phi(r,\theta)}
\nonumber \\
J_{\scriptscriptstyle{L}} & = &
(2 \pi\,a^{2L+1})^{-1}\,
\int_0^a dr\,r^{2L}
\int_0^{2 \pi}d \theta\,e^{i \theta}\,e^{2 \phi(r,\theta)}\,.
\label{eqn:ivthirtyone}
\end{eqnarray}
Similarly, (\ref{eqn:ivtwentyfour}), (\ref{eqn:ivtwentyeight}), 
(\ref{eqn:ivtwentynine}) and (\ref{eqn:ivseventeen}) with 
$\Delta_{\scriptscriptstyle{L}}$ replaced with $\Delta_{\scriptscriptstyle{L+1}}$ give
\begin{eqnarray}
& &i ka\,
(e^{-2 i ka} +e^{ 2 i \Delta_{\scriptscriptstyle{L+1}}}\,
\cos 2 \epsilon )\,I_{\scriptscriptstyle{L+1}}= \nonumber \\
&&e^{ 2 i \Delta_{\scriptscriptstyle{L+1}}}\,\cos 2 \epsilon \, - e^{-2 i ka}
 (1- i ka)\,J_{\scriptscriptstyle{L}}^{\star}\,
e^{i (\Delta_{\scriptscriptstyle{L}} +\Delta_{\scriptscriptstyle{L+1}} + \lambda - \pi L)}\,
\sin 2 \epsilon
+ \mbox{O}((ka)^3, (ka)^2\Delta_{\scriptscriptstyle{L+1}})\,.
\label{eqn:ivthirtytwo}
\end{eqnarray}
Equations (\ref{eqn:ivthirty}) and (\ref{eqn:ivthirtytwo}) can be solved for
$\Delta_{\scriptscriptstyle{L}}$, $\Delta_{\scriptscriptstyle{L+1}}$, $\epsilon$ and
 $\lambda$ with
\begin{equation}
\lambda = \lambda_0 + \lambda_1 k a + \mbox{O}((ka)^2)\,.
\label{eqn:ivthirtythree}
\end{equation}
The results are
\begin{eqnarray}
\Delta_{\scriptscriptstyle{L}} & = &
- ka/( 1 + I_{\scriptscriptstyle{L}}) +  \mbox{O}((ka)^3) \nonumber \\
\Delta_{\scriptscriptstyle{L+1}} & = &
[ I_{\scriptscriptstyle{L+1}} - 1 - 
(\cos \lambda_0 \mbox{Im}J_{\scriptscriptstyle{L}}- 
\sin \lambda_0 \mbox{Re}J_{\scriptscriptstyle{L}})^2/( 1 + I_{\scriptscriptstyle{L}})
]\, ka + \mbox{O}((ka)^3) \nonumber \\
0 &=&\cos \lambda_0 \mbox{Re}J_{\scriptscriptstyle{L}} +
\sin \lambda_0 \mbox{Im}J_{\scriptscriptstyle{L}} \nonumber \\
\lambda_1& = &0 \nonumber \\
\epsilon &=& (-1)^L\,
(\sin \lambda_0 \mbox{Re}J_{\scriptscriptstyle{L}} - 
\cos \lambda_0 \mbox{Im}J_{\scriptscriptstyle{L}})\,ka/( 1 + I_{\scriptscriptstyle{L}})
+  \mbox{O}((ka)^2)\,.
\label{eqn:ivthirtyfour}
\end{eqnarray}
Note that $\Delta_{\scriptscriptstyle{L}}$ and $\Delta_{\scriptscriptstyle{L+1}}$ have the 
same energy dependence as in the centrally symmetric case given by
(\ref{eqn:ivtwo}).
\renewcommand{\thesection}{\Roman{section}}
\renewcommand{\theequation}{\thesection.\arabic{equation}}
\section{{\bf CHIRAL LIMIT}}
\setcounter{equation}{0}
\renewcommand{\thesection}{\arabic{section}}
\renewcommand{\theequation}{\thesection.\arabic{equation}}

We concluded in Sec. IV that general magnetic fields only modify the numerical coefficients of the energy-dependent part of the phase shifts calculated in the
centrally symmetric case. It was also noted that the off-diagonal elements of
$S_{\scriptscriptstyle{l,L}}$ fall off as powers of $k$. Therefore, inhomogenous fields of the type considered here do not result in special cases that are not already included in the  central symmetry limit. We therefore confine our 
discussion to  central 
symmetry from here on and proceed to show that (\ref{eqn:ione}) is true by
demonstrating that the integral in (\ref{eqn:iitwelve}) satisfies
\begin{equation}
\lim_{m^2 \to 0}
m^2
\int_0^{\infty}dt\,e^{-t m^2}
\int d^2 r
\int_0^M dk\,k \,e^{-k^2 t}
\int_0^{2 \pi}d \Theta\,
( | \psi^{(+)}({\bf k, r})|^2 -  |\psi_0({\bf k, r})|^2 )
= 0\,.
\label{eqn:vone}
\end{equation}
This integral can be divided between contributions from the regions $r < a$ and $r > a$. We may set $a=R$. 
\subsection{{\bf ${\bf r < a}$ }}
By inspection, the term $|\psi_0|^2$ in (\ref{eqn:vone}) gives a contribution
proportional to $\ln [(M^2 + m^2)/m^2]$ and so vanishes in the indicated
limit. This leaves an integral over partial waves obtained by substituting
(\ref{eqn:iiione}) in (\ref{eqn:vone}). We can obtain a bound on 
$\psi_{\scriptscriptstyle{l}}(k, r)$ for all $l$ from (\ref{eqn:iiitwentysix}) and 
(\ref{eqn:iiifivtytwo}) with $b_{\scriptscriptstyle{l-1}}$,  $g_{\scriptscriptstyle{-l}}$ 
fixed by (\ref{eqn:iiifortysix})-(\ref{eqn:iiifortyseven}) and 
 (\ref{eqn:iiifortynine}), respectively. These equations and the estimates
\begin{eqnarray}
r^{-l}\,e^{-\phi(r)}\,
\int_0^r dx\,x^{2l-1}\,e^{2 \phi(x)}
& \leq &
\frac{a^l}{2 l}\, e^{3 a ||B||/\sqrt2 \pi}\,,
\nonumber \\
\int_0^a dx\,x^{2l-1}\,e^{2 \phi(x)}
& \geq &
a^{2l}\,e^{- a ||B||/ \sqrt \pi}/2l\,,
\label{eqn:vtwo}
\end{eqnarray}
following from (\ref{eqn:iiifivtysix}) give
\begin{equation}
|\psi_{\scriptscriptstyle{l}}(k, r)|
\leq
\frac{ e^{5 a ||B||/\sqrt2 \pi} }{ \sqrt 2 \Gamma(W+1)}\,
\left( \frac{ka}{2} \right)^W\, ( 1+ \mbox{O} (k^2) )\,,
\label{eqn:vthree}
\end{equation}
where the $ \mbox{O} (k^2)$ term symbolizes a remainder term that vanishes
as $k \rightarrow 0$ for all $l$ and that falls off as $k^2/|l|$ for
$|l| \gg \Phi/2 \pi$. Recall that $W = | l - \Phi/2 \pi|$.
From (\ref{eqn:vthree}),
\begin{eqnarray}
\int_0^{\infty}dt\,e^{-t m^2}
\int_0^a dr\,r & &
\int_0^M dk\,k \,e^{-k^2 t}
\sum_l |\psi_{\scriptscriptstyle{l}}(k, r)|^2 \nonumber \\
\leq
& &
\frac{a^2}{8}\, e^{5 a ||B||/\sqrt\pi}\, 
\ln \left( \frac{M^2 + m^2}{m^2} \right)\,
\sum_l \frac{ (M a/2)^{2W}}{\Gamma^2(W+1)}\,,
\label{eqn:vfour}
\end{eqnarray}
for $ M a \ll 1$, and hence the indicated limit in  (\ref{eqn:vone}) is
satisfied. The special case when $l = \Phi/ 2 \pi= 1, 2, \ldots$ is dealt
with by (\ref{eqn:iiifortyeight}) and (\ref{eqn:ivseven}). For these special
values of $l$ the integrals on the left-hand side of (\ref{eqn:vfour}) 
contribute a term of order $M^2 a^2 [(M^2 + m^2) \ln^2 (M a)]^{-1}$,
which vanishes in the limit indicated in (\ref{eqn:vone})
\subsection{{\bf ${\bf r > a}$}}
Equations (\ref{eqn:iiione}), (\ref{eqn:iiifortythree}) and the expansion 
(\ref{eqn:iiifive}) substituted into (\ref{eqn:vone}) result in the following
integral:
\begin{eqnarray}
I =
\int_0^{\infty}dt\,e^{-t m^2}
\int_a^{\infty} dr\,r & & 
\int_0^M dk\,k \,e^{-k^2 t}
\sum_l 
\left[
J_{\scriptscriptstyle{W}}^2(kr) - J_{\scriptscriptstyle{l}}^2(kr) \right.
\nonumber \\
& &
+ 
\left.
J_{\scriptscriptstyle{W}}(kr) \, Y_{\scriptscriptstyle{W}}(kr)\,
\sin 2 \Delta_{\scriptscriptstyle{l}}
+
( Y_{\scriptscriptstyle{W}}^2(kr) - J_{\scriptscriptstyle{W}}^2(kr) )\,
\sin^2 \Delta_{\scriptscriptstyle{l}}
\right]\,.
\label{eqn:vfive}
\end{eqnarray}
Consider the sum over the first two Bessel functions.
Entries 5.7.11.6 of Ref.20 and 6.538.1 of Ref.21
give the result
\begin{eqnarray}
\sum_l
[ J_{\scriptscriptstyle{W}}^2(kr) - J_{\scriptscriptstyle{l}}^2(kr)]
& = &
1/2\,
J_{\scriptscriptstyle{f}}^2(kr)
+ 
1/2\,
J_{\scriptscriptstyle{1-f}}^2(kr)
\nonumber \\
&- &
\int_{kr}^{\infty} dt\, t^{-1}\,
[ f \,J_{\scriptscriptstyle{f}}^2(t)
+ (1-f)\,J_{\scriptscriptstyle{1-f}}^2(t) ]\,. \nonumber \\
&\equiv &
g(k r)\,,
\label{eqn:vsix}
\end{eqnarray}
where $ \Phi/ 2 \pi = N + f$, $ 0 < f < 1$ and $N = 0, 1,..$. 
Next (\ref{eqn:vsix}) has to be integrated over $k$ following (\ref{eqn:vfive}). For this we apply the weighted mean value theorem~\cite{xi}:
Assume $f$ and $g$ are continuous on $[a, b]$. If $f$ never changes sign on
$[a, b]$ then, for some $c$ in $[a, b]$,
$$
\int_a^b\,
f(k)\,g(k)\,dk =
g(c)\,
\int_a^b\,
f(k)\,dk\,.
$$
Let $f = k e^{- k^2 t}$ and $g$ equal the right-hand side of (\ref{eqn:vsix}).
Then for some $\mu$ satisfying $0 < \mu \le M$,
\begin{equation}
\int_0^M dk\, k\, e^{- k^2 t}
\sum_l 
(J_{\scriptscriptstyle{W}}^2 - J_{\scriptscriptstyle{l}}^2 )
=
( 1 - e^{ -M^2 t})\, g( \mu r)/2t\,.
\label{eqn:vseven}
\end{equation}
The value $\mu = 0$ is excluded since the energy integral is manifestly 
$r$- dependent. The $t$-integral in (\ref{eqn:vfive}) can be done immediately,
resulting in an overall factor of $\ln [ (M^2 + m^2)/m^2]$.
\paragraph*{}
It remains to be shown that the integration over $r$ is bounded. The definition  of
$g$ in (\ref{eqn:vsix}) for large argument gives
\begin{equation}
g(z)
=
- \frac{ \sin \pi f\, \cos 2 z}{ \pi z}
+
\frac{ (f^2 -f - \frac{1}{4})\,  \sin \pi f\, \sin 2 z}{  \pi z^2}
+
\mbox{O}
\left(
\frac{ \cos 2 z,  \sin 2 z}{z^3}
\right)\,.
\label{eqn:veight}
\end{equation}
Subsitution of (\ref{eqn:veight}) into (\ref{eqn:vseven}) and performing the 
$r$-integral in (\ref{eqn:vfive}) results in
\begin{eqnarray}
\int_0^{\infty}dt\,e^{-t m^2}
& &
\lim_{L \to \infty}
\int_a^L dr\,r 
\int_0^M dk\,k \,e^{-k^2 t}
\sum_l [J_{\scriptscriptstyle{W}}^2 - J_{\scriptscriptstyle{l}}^2] \nonumber \\
& &=
\ln \left( \frac{M^2 + m^2}{m^2} \right)\,
\lim_{L \to \infty}
\left[
-\frac{ \sin \pi f}{4 \pi \mu^2}\, \sin 2 \mu L +
\mbox{convergent as}\, L \rightarrow \infty
\right]\,.
\label{eqn:vnine}
\end{eqnarray}
The leading term, although oscillating, is bounded and that is all that is 
required to satisfy the limit indicated in  (\ref{eqn:vone}). For the 
special case when $ \Phi/ 2 \pi = 1, 2,..$ the sum in  (\ref{eqn:vsix}) 
is zero.
\paragraph*{}
Next we consider the $J_{\scriptscriptstyle{W}}\,Y_{\scriptscriptstyle{W}}$ terms in
(\ref{eqn:vfive}). Since $J_{\scriptscriptstyle{W}}^2(kz) +Y_{\scriptscriptstyle{W}}^2(kz)$ 
is a decreasing function of $z$ for any value of $W$~\cite{xvi}, then for 
$r \geq a$,
\begin{equation}
Y_{\scriptscriptstyle{W}}^2(kr) \leq
J_{\scriptscriptstyle{W}}^2(ka) + Y_{\scriptscriptstyle{W}}^2(ka)\,.
\label{eqn:vten}
\end{equation}
It follows that
\begin{equation}
\left|
\sum_l
J_{\scriptscriptstyle{W}}(kr)\,Y_{\scriptscriptstyle{W}}(kr)
\sin (2 \Delta_{\scriptscriptstyle{l}})
\right|
\leq
\sum_l
\left|
\sin (2 \Delta_{\scriptscriptstyle{l}})
\right|\,
\,[ 1 + Y_{\scriptscriptstyle{W}}^2(ka)]^{1/2}\,,
\label{eqn:veleven}
\end{equation}
where we used $ |J_{\scriptscriptstyle{W}}(z)| \leq 1$ for $ W \geq 0$. 
From  (\ref{eqn:ivtwo}) together with  (\ref{eqn:ivfive}),  (\ref{eqn:ivsix}) 
and  (\ref{eqn:iiifivtysix}) one obtains for all $l$ and 
$\Phi/ 2 \pi \neq 1, 2, \ldots$
\begin{equation}
|\Delta_{\scriptscriptstyle{l}}|
\leq
\pi \, 
\left(
e^{a ||B||/ \sqrt \pi} + 1
\right)\,
\frac{ (ka/2)^{2W}}{\Gamma(W)\, \Gamma(W+1)}\,
(1 + \mbox{O}(k^2) )\,,
\label{eqn:vtwelve}
\end{equation}
where the $\mbox{O}(k^2)$ terms fall off for large $|l|$ at least like 
$1/|l|$. Since $ka \ll 1$ and
$$
Y_{\scriptscriptstyle{W}}(ka)
\sim
- \frac{1}{\pi}\, \Gamma(W)\, (ka/2)^{-W}\,,
$$
every term in the series on the left-hand side of (\ref{eqn:veleven}) is 
bounded by a constant
$$
2^{3/2}\,
\left(
e^{a ||B||/ \sqrt \pi} + 1
\right)\,
\frac{ (M a/2)^{W}}{\Gamma(W+1)}\,,
$$
for all $r > a$ and $ 0 \leq k \leq M$. It is, therefore, a uniformly 
covergent series of continuous functions of $k$ and can be integrated term by 
term. Applying the weighted mean value theorem again we obtain
\begin{eqnarray}
\int_0^{\infty}dt\,e^{-t m^2}
& &
\lim_{L \to \infty}
\int_a^L dr\,r 
\int_0^M dk\,k \,
\sum_l J_{\scriptscriptstyle{W}}(kr)\,Y_{\scriptscriptstyle{W}}(kr)\,
\sin 2 \Delta_{\scriptscriptstyle{l}}(k) \nonumber \\
& &=
\frac{1}{2}\,\ln \left( \frac{M^2 + m^2}{m^2} \right)\,
\lim_{L \to \infty}
\int_a^L dr\,r 
\sum_l J_{\scriptscriptstyle{W}}(\mu r)\,Y_{\scriptscriptstyle{W}}(\mu r)\,
\sin 2 \Delta_{\scriptscriptstyle{l}}(\mu)\,,
\label{eqn:vthirteen}
\end{eqnarray}
for some $\mu$ in the interval $ 0 < \mu \leq M$. For $|l| \gg \Phi/2 \pi$, 
$ J_{\scriptscriptstyle{W}}(\mu r)\,Y_{\scriptscriptstyle{W}}(\mu r) \sim 
- ( \pi |l|)^{-1}$ which, together with (\ref{eqn:vtwelve}), implies each term 
in the series on the right-hand side of (\ref{eqn:vthirteen}) is dominated by a
 constant whose $l$-dependence is $( \mu a /2 )^{2 |l|}/(|l!|)^2$ for 
all $r > a$. For all finite $L > a$ it is a uniformly convergent series of 
continuous functions of $r$ that can be integrated term by term. From entry 
5.11.10 of Ref.23,
\begin{eqnarray}
\int_a^L dr\,r\,J_{\scriptscriptstyle{W}}(\mu r)\,Y_{\scriptscriptstyle{W}}(\mu r)
& = &
\frac{1}{4}\, L^2
\left[
2 J_{\scriptscriptstyle{W}}(\mu L)\,Y_{\scriptscriptstyle{W}}(\mu L) -
J_{\scriptscriptstyle{W-1}}(\mu L)\,Y_{\scriptscriptstyle{W+1}}(\mu L)
\right. \nonumber \\
& &
\left.
- J_{\scriptscriptstyle{W+1}}(\mu L)\,Y_{\scriptscriptstyle{W-1}}(\mu L)
\right]
-(L \rightarrow a) \nonumber \\
& \equiv & 
h_{\scriptscriptstyle{l}}(L) - h_{\scriptscriptstyle{l}}(a)\,.
\label{eqn:vfourteen}
\end{eqnarray}
There remains 
$
\lower5pt \hbox{$\scriptscriptstyle L \to \infty $}
\mkern-30mu \lim \;\;
\lower8pt \hbox{$\scriptscriptstyle l$}
\mkern-15mu \sum
h_{\scriptscriptstyle{l}}(L)\,\sin 2 \Delta_{\scriptscriptstyle{l}}(\mu)\,.
$
For $|l| \gg \Phi/2 \pi$, 
\begin{equation}
h_{\scriptscriptstyle{l}}(L)
=
\frac{|l|}{\pi \mu^2} + \mbox{O} \left( \frac{1}{|l|}\right)\,,
\label{eqn:vfivteen}
\end{equation}
which, together with (\ref{eqn:vtwelve}), implies 
$h_{\scriptscriptstyle{l}}(L)\,\sin 2 \Delta_{\scriptscriptstyle{l}}(\mu)$ is dominated by 
a term whose $l$-dependence is $( \mu a /2 )^{2 |l|}/[ (|l| -1)! ]^2$ for all 
finite $L$ and $ \mu a \ll 1$. Therefore the series
$ 
\lower8pt \hbox{$\scriptscriptstyle l$}
\mkern-15mu \sum 
h_{\scriptscriptstyle{l}}(L)\,\sin 2 \Delta_{\scriptscriptstyle{l}}$ is uniformly 
and absolutely convergent for all finite values of $ L > a$. But it does not 
necessarily converge to a function continuous at the point $L = \infty$ since 
$h_{\scriptscriptstyle{l}}(L)$ is not continuous at $L = \infty$. 
In fact, for fixed $l$ and $\mu L \gg 1$,
\begin{equation}
h_{\scriptscriptstyle{l}}(L)=
-\frac{ \sin(2 \mu L - \pi W)}{2 \pi \mu^2} + 
\mbox{O}\left( \frac{1}{L}\right)\,.
\label{eqn:vsixteen}
\end{equation}
The remedy is now clear. Consider instead the two series
\begin{eqnarray}
\sum_l
\left[
h_{\scriptscriptstyle{l}}(L)
+ 
\frac{\sin(2 \mu L - \pi W)}{2 \pi \mu^2} 
\right] & 
\sin 2 \Delta_{\scriptscriptstyle{l}} &\nonumber \\
&- & \frac{1}{2 \pi \mu^2}
\sum_l \sin(2\Delta_{\scriptscriptstyle{l}})\,\sin(2 \mu L - \pi W)\,. \nonumber
\end{eqnarray}
Now the limit $L \rightarrow \infty$ and the sum can be interchanged in the first series, giving zero. The second series is bounded for all $L > a$, and so 
the limit in (\ref{eqn:vone}) is true for the 
$J_{\scriptscriptstyle{W}}\,Y_{\scriptscriptstyle{W}}$ terms.
\paragraph*{}
The special case $ \Phi/ 2 \pi= 1, 2, \ldots$ requires (\ref{eqn:ivseven}) 
when $l =  \Phi/ 2 \pi$. This results in a term 
$J_{\scriptscriptstyle{0}}(kr)\,Y_{\scriptscriptstyle{0}}(kr)/\ln(ka)$, which is 
continuous for $k$ on $[0,M]$. Therefore the weighted mean value theorem may 
be applied again to the $k$-integral, resulting in an overall factor of 
$\ln[(M^2 +m^2)/m^2]$. There remains an integration over 
$r\,J_{\scriptscriptstyle{0}}(\mu r)\,Y_{\scriptscriptstyle{0}}(\mu r)$ between $a$ and 
$L$, giving an oscillating but bounded term $\sin (2 \mu L)$ as
$L \rightarrow \infty$. Again, (\ref{eqn:vone}) is satisfied for this special 
term.
\paragraph*{}
Finally, consider the last two terms in (\ref{eqn:vfive}). Note that
\begin{equation}
|Y_{\scriptscriptstyle{W}}^2(kr) - J_{\scriptscriptstyle{W}}^2(kr)|
\leq
Y_{\scriptscriptstyle{W}}^2(ka) + J_{\scriptscriptstyle{W}}^2(ka)\,,
\label{eqn:vseventeen}
\end{equation}
for $r \geq a$ since $J_{\scriptscriptstyle{W}}^2(z)+ Y_{\scriptscriptstyle{W}}^2(z)$ is a 
decreasing function of $z$~\cite{xvi}. Hence, for $ka \ll 1$, 
$ 0 \leq k \leq M$,
\begin{equation}
|Y_{\scriptscriptstyle{W}}^2(kr) - J_{\scriptscriptstyle{W}}^2(kr)|
\,\sin^2 \Delta_{\scriptscriptstyle{l}}
\leq
\left(
e^{a ||B||/ \sqrt \pi} + 1
\right)^2\,
\frac{ (M a/2)^{2W}}{\Gamma^2(W+1)}\,
(1 + \mbox{O}(  (Ma)^2,(Ma)^{2W} ) )\,,
\label{eqn:veighteen}
\end{equation}
so that the sum over $l$ of the terms on the left-hand side of  
(\ref{eqn:veighteen}) converges uniformly for all $r \geq a$, 
$0 \leq k \leq M$. As it is also a sum of continuous functions of $k$ for
 $0 \leq k \leq M$ it can be integrated term by term over $k$. Application of 
the weighted mean value theorem gives
\begin{eqnarray}
& &
\int_0^{\infty}dt\,e^{-t m^2}
\lim_{L \to \infty}
\int_a^L dr\,r 
\int_0^M dk\,k
\sum_l [Y_{\scriptscriptstyle{W}}^2(kr) - J_{\scriptscriptstyle{W}}^2(kr)] \,
\sin^2 \Delta_{\scriptscriptstyle{l}}(k)
\nonumber \\
& &=
\frac{1}{2}\,
\ln \left( \frac{M^2 + m^2}{m^2} \right)\,
\lim_{L \to \infty}
\int_a^L dr\,r
\sum_l [Y_{\scriptscriptstyle{W}}^2(\mu r) - J_{\scriptscriptstyle{W}}^2(\mu r)] \,
\sin^2 \Delta_{\scriptscriptstyle{l}}(\mu)\,,
\label{eqn:vnineteen}
\end{eqnarray}
for some $\mu$ in the interval  $0 <  \mu \leq M$. For $|l| \gg \Phi/2 \pi$ 
each term in the series (\ref{eqn:vnineteen}) is dominated by a $r$-
independent constant whose $l$-dependence is $(\mu a/2)^{2 |l|}/( |l!| )^2$, 
for $ r \geq a$ and $ \mu a \ll 1$. It is therefore a uniformly convergent 
series of continuous functions for all finite $L > a$ that can be integrated 
term by term.
Entry 5.54.2 in Ref.21 gives
\begin{eqnarray}
\int_a^L dr\,r \,
[Y_{\scriptscriptstyle{W}}^2(\mu r) - J_{\scriptscriptstyle{W}}^2(\mu r)]
& = &
\frac{L^2}{2}\,
\left[
Y_{\scriptscriptstyle{W}}^2(\mu L) - 
Y_{\scriptscriptstyle{W-1}}(\mu L)\,Y_{\scriptscriptstyle{W+1}}(\mu L)
\right. \nonumber \\
& - &
\left.
J_{\scriptscriptstyle{W}}^2(\mu L) + 
J_{\scriptscriptstyle{W+1}}(\mu L)\,J_{\scriptscriptstyle{W-1}}(\mu L)
\right] - (L \rightarrow a) \nonumber \\
& \equiv& 
k_{\scriptscriptstyle{l}}(L) - k_{\scriptscriptstyle{l}}(a)\,.
\label{eqn:vtwenty}
\end{eqnarray}
Next, consider 
$
\lower5pt \hbox{$\scriptscriptstyle L \to \infty $}
\mkern-30mu \lim \;\;
\lower8pt \hbox{$\scriptscriptstyle l$}
\mkern-15mu \sum
k_{\scriptscriptstyle{l}}(L)\,\sin^2 \Delta_{\scriptscriptstyle{l}}(\mu)
$. For $ \mu L \gg 1$,
\begin{equation}
k_{\scriptscriptstyle{l}}(L)=
\frac{ \cos( 2 \mu L - \pi W)}{ \pi \mu^2}  +
\mbox{O} \left( \frac{1}{L}\right)\,,
\label{eqn:vtwentyone}
\end{equation}
and hence consider the series
\begin{eqnarray}
\sum_l
\left[
k_{\scriptscriptstyle{l}}(L)
- \frac{ \cos( 2 \mu L - \pi W)}{ \pi \mu^2} 
\right]& 
\sin^2 \Delta_{\scriptscriptstyle{l}} &\nonumber \\
&+&  \frac{1}{\pi \mu^2}\,
\sum_l
\cos( 2 \mu L - \pi W)\,\sin^2 \Delta_{\scriptscriptstyle{l}}\,. \nonumber
\end{eqnarray}
The first series is a sum of continuous functions for all $L > a$. It is also 
a uniformly and absolutely convergent series for $L > a$, first because 
$|J_{\scriptscriptstyle{W}}| \leq 1$ and, secondly, the combinations 
$Y_{\scriptscriptstyle{W}}^2\,\sin^2 \Delta_{\scriptscriptstyle{l}}$ and 
$Y_{\scriptscriptstyle{W-1}}\,Y_{\scriptscriptstyle{W+1}}\,
\sin^2 \Delta_{\scriptscriptstyle{l}}$ are dominated by $L$-independent constants 
whose $l$-dependence is $( \mu a\/2)^{2 |l|}/(|l|!)^2$ for $|l| \gg \Phi/2 \pi$. 
Hence the limit $L \rightarrow \infty$ and the sum can be interchanged in the 
first series, giving zero. The second series is bounded for all $ L > a$, 
verifying (\ref{eqn:vone}) for the last series of terms in (\ref{eqn:vfive}).
\paragraph*{}
The special case $\Phi/2 \pi = 1, 2 \ldots$ is dealt with in the same way 
as in the case of 
$J_{\scriptscriptstyle{W}}\,Y_{\scriptscriptstyle{W}}\, \sin \Delta_{\scriptscriptstyle{l}}$ and 
gives a contribution that vanishes in the limit indicated in (\ref{eqn:vone}).
Thus, (\ref{eqn:ione}) is demonstrated.
\acknowledgments
The author has benefited from discussions with C.~Adam concerning the 
relation between massive $\mbox{QED}_2$ and the Schwinger model. 
He also gratefully acknowledges support from a Forbairt Basic Research Grant.

\end{document}